\documentclass[longauth]{aa}  

\usepackage{color}

\newcommand{\Ni}{\ensuremath{^{56}\mathrm{Ni}}}

\newcommand{\Msun}{\ensuremath{\mathrm{M}_\odot}}

\graphicspath{{./}{figures/}}

\usepackage{graphicx}
\usepackage{txfonts}
\usepackage[]{hyperref}
\begin{document}

   \title{\textit{Euclid}: Searching for pair-instability supernovae with the Deep Survey\thanks{This paper is published on behalf of the Euclid Consortium.}}


\newcommand{\orcid}[1]{} 
\author{T.J.~Moriya\orcid{0000-0003-1169-1954}$^{1,2}$\thanks{\email{takashi.moriya@nao.ac.jp}}, C.~Inserra\orcid{0000-0002-3968-4409}$^{3}$, M.~Tanaka$^{4,5}$, E.~Cappellaro\orcid{0000-0001-5008-8619}$^{6}$, M.~Della Valle\orcid{0000-0003-3142-5020}$^{7,8,9}$, I.~Hook\orcid{0000-0002-2960-978X}$^{10}$, R.~Kotak\orcid{0000-0001-5455-3653}$^{11}$, G.~Longo\orcid{0000-0002-9182-8414}$^{12,8}$, F.~Mannucci\orcid{0000-0002-4803-2381}$^{13}$, S.~Mattila\orcid{0000-0001-7497-2994}$^{11}$
% ...
}

\institute{$^{1}$ National Astronomical Observatory of Japan, National Institutes of Natural Sciences, 2-21-1 Osawa, Mitaka, Tokyo 181-8588, Japan\\
$^{2}$ School of Physics and Astronomy, Faculty of Science, Monash University, Clayton, Victoria 3800, Australia\\
$^{3}$ School of Physics and Astronomy, Cardiff University, The Parade, Cardiff, CF24 3AA, UK\\
$^{4}$ Astronomical Institute, Tohoku University, Sendai 980-8578, Japan\\
$^{5}$ Division for the Establishment of Frontier Sciences, Organization for Advanced Studies, Tohoku University, Sendai 980-8577, Japan\\
$^{6}$ INAF-Osservatorio Astronomico di Padova, Via dell'Osservatorio 5, I-35122 Padova, Italy\\
$^{7}$ INAF-Osservatorio Astronomico di Capodimonte, Via Moiariello 16, I-80131 Napoli, Italy\\
$^{8}$ INFN section of Naples, Via Cinthia 6, I-80126, Napoli, Italy\\
$^{9}$ International Center for Relativistic Astrophysics, Piazzale della Republica 2, I-65122 Pescara, Italy\\
$^{10}$ Department of Physics, Lancaster University, Lancaster, LA1 4YB, UK\\
$^{11}$ Department of Physics and Astronomy, Vesilinnantie 5, 20014 University of Turku, Finland\\
$^{12}$ Department of Physics "E. Pancini", University Federico II, Via Cinthia 6, I-80126, Napoli, Italy\\
$^{13}$ INAF-Osservatorio Astrofisico di Arcetri, Largo E. Fermi 5, I-50125, Firenze, Italy\\
$^{14}$ Aix-Marseille Univ, CNRS/IN2P3, CPPM, Marseille, France\\
$^{15}$ ESAC/ESA, Camino Bajo del Castillo, s/n., Urb. Villafranca del Castillo, 28692 Villanueva de la Ca\~nada, Madrid, Spain\\
$^{16}$ Institute of Cosmology and Gravitation, University of Portsmouth, Portsmouth PO1 3FX, UK\\
$^{17}$ INAF-Osservatorio di Astrofisica e Scienza dello Spazio di Bologna, Via Piero Gobetti 93/3, I-40129 Bologna, Italy\\
$^{18}$ INAF-Osservatorio Astrofisico di Torino, Via Osservatorio 20, I-10025 Pino Torinese (TO), Italy\\
$^{19}$ INFN-Sezione di Roma Tre, Via della Vasca Navale 84, I-00146, Roma, Italy\\
$^{20}$ Dipartimento di Fisica, Universit\'a degli studi di Genova, and INFN-Sezione di Genova, via Dodecaneso 33, I-16146, Genova, Italy\\
$^{21}$ Instituto de Astrof\'isica e Ci\^encias do Espa\c{c}o, Universidade do Porto, CAUP, Rua das Estrelas, PT4150-762 Porto, Portugal\\
$^{22}$ Dipartimento di Fisica, Universit\'a degli Studi di Torino, Via P. Giuria 1, I-10125 Torino, Italy\\
$^{23}$ INFN-Sezione di Torino, Via P. Giuria 1, I-10125 Torino, Italy\\
$^{24}$ INAF-IASF Milano, Via Alfonso Corti 12, I-20133 Milano, Italy\\
$^{25}$ Port d'Informaci\'{o} Cient\'{i}fica, Campus UAB, C. Albareda s/n, 08193 Bellaterra (Barcelona), Spain\\
$^{26}$ Institut de F\'{i}sica d'Altes Energies (IFAE), The Barcelona Institute of Science and Technology, Campus UAB, 08193 Bellaterra (Barcelona), Spain\\
$^{27}$ INAF-Osservatorio Astronomico di Roma, Via Frascati 33, I-00078 Monteporzio Catone, Italy\\
$^{28}$ Dipartimento di Fisica e Astronomia "Augusto Righi" - Alma Mater Studiorum Universit\`{a} di Bologna, via Piero Gobetti 93/2, I-40129 Bologna, Italy\\
$^{29}$ Institut national de physique nucl\'eaire et de physique des particules, 3 rue Michel-Ange, 75794 Paris C\'edex 16, France\\
$^{30}$ Centre National d'Etudes Spatiales, Toulouse, France\\
$^{31}$ Institute for Astronomy, University of Edinburgh, Royal Observatory, Blackford Hill, Edinburgh EH9 3HJ, UK\\
$^{32}$ Jodrell Bank Centre for Astrophysics, Department of Physics and Astronomy, University of Manchester, Oxford Road, Manchester M13 9PL, UK\\
$^{33}$ European Space Agency/ESRIN, Largo Galileo Galilei 1, 00044 Frascati, Roma, Italy\\
$^{34}$ Univ Lyon, Univ Claude Bernard Lyon 1, CNRS/IN2P3, IP2I Lyon, UMR 5822, F-69622, Villeurbanne, France\\
$^{35}$ Institute of Physics, Laboratory of Astrophysics, Ecole Polytechnique F\'{e}d\'{e}rale de Lausanne (EPFL), Observatoire de Sauverny, 1290 Versoix, Switzerland\\
$^{36}$ Mullard Space Science Laboratory, University College London, Holmbury St Mary, Dorking, Surrey RH5 6NT, UK\\
$^{37}$ Departamento de F\'isica, Faculdade de Ci\^encias, Universidade de Lisboa, Edif\'icio C8, Campo Grande, PT1749-016 Lisboa, Portugal\\
$^{38}$ Instituto de Astrof\'isica e Ci\^encias do Espa\c{c}o, Faculdade de Ci\^encias, Universidade de Lisboa, Campo Grande, PT-1749-016 Lisboa, Portugal\\
$^{39}$ Department of Astronomy, University of Geneva, ch. d\'Ecogia 16, CH-1290 Versoix, Switzerland\\
$^{40}$ Universit\'e Paris-Saclay, CNRS, Institut d'astrophysique spatiale, 91405, Orsay, France\\
$^{41}$ Department of Physics, Oxford University, Keble Road, Oxford OX1 3RH, UK\\
$^{42}$ INFN-Padova, Via Marzolo 8, I-35131 Padova, Italy\\
$^{43}$ Universit\'e Paris-Saclay, Universit\'e Paris Cit\'e, CEA, CNRS, Astrophysique, Instrumentation et Mod\'elisation Paris-Saclay, 91191 Gif-sur-Yvette, France\\
$^{44}$ INAF-Osservatorio Astronomico di Trieste, Via G. B. Tiepolo 11, I-34143 Trieste, Italy\\
$^{45}$ Istituto Nazionale di Astrofisica (INAF) - Osservatorio di Astrofisica e Scienza dello Spazio (OAS), Via Gobetti 93/3, I-40127 Bologna, Italy\\
$^{46}$ Istituto Nazionale di Fisica Nucleare, Sezione di Bologna, Via Irnerio 46, I-40126 Bologna, Italy\\
$^{47}$ Max Planck Institute for Extraterrestrial Physics, Giessenbachstr. 1, D-85748 Garching, Germany\\
$^{48}$ Universit\"ats-Sternwarte M\"unchen, Fakult\"at f\"ur Physik, Ludwig-Maximilians-Universit\"at M\"unchen, Scheinerstrasse 1, 81679 M\"unchen, Germany\\
$^{49}$ Institute of Theoretical Astrophysics, University of Oslo, P.O. Box 1029 Blindern, N-0315 Oslo, Norway\\
$^{50}$ Jet Propulsion Laboratory, California Institute of Technology, 4800 Oak Grove Drive, Pasadena, CA, 91109, USA\\
$^{51}$ von Hoerner \& Sulger GmbH, Schlo{\ss}Platz 8, D-68723 Schwetzingen, Germany\\
$^{52}$ Technical University of Denmark, Elektrovej 327, 2800 Kgs. Lyngby, Denmark\\
$^{53}$ Max-Planck-Institut f\"ur Astronomie, K\"onigstuhl 17, D-69117 Heidelberg, Germany\\
$^{54}$ Department of Physics and Helsinki Institute of Physics, Gustaf H\"allstr\"omin katu 2, 00014 University of Helsinki, Finland\\
$^{55}$ NOVA optical infrared instrumentation group at ASTRON, Oude Hoogeveensedijk 4, 7991PD, Dwingeloo, The Netherlands\\
$^{56}$ Argelander-Institut f\"ur Astronomie, Universit\"at Bonn, Auf dem H\"ugel 71, 53121 Bonn, Germany\\
$^{57}$ INFN-Sezione di Bologna, Viale Berti Pichat 6/2, I-40127 Bologna, Italy\\
$^{58}$ Department of Physics, Institute for Computational Cosmology, Durham University, South Road, DH1 3LE, UK\\
$^{59}$ Institut d'Astrophysique de Paris, UMR 7095, CNRS, and Sorbonne Universit\'e, 98 bis boulevard Arago, 75014 Paris, France\\
$^{60}$ University of Applied Sciences and Arts of Northwestern Switzerland, School of Engineering, 5210 Windisch, Switzerland\\
$^{61}$ European Space Agency/ESTEC, Keplerlaan 1, 2201 AZ Noordwijk, The Netherlands\\
$^{62}$ Department of Physics and Astronomy, University of Aarhus, Ny Munkegade 120, DK-8000 Aarhus C, Denmark\\
$^{63}$ Institute of Space Science, Bucharest, Ro-077125, Romania\\
$^{64}$ IFPU, Institute for Fundamental Physics of the Universe, via Beirut 2, 34151 Trieste, Italy\\
$^{65}$ Dipartimento di Fisica e Astronomia "G.Galilei", Universit\'a di Padova, Via Marzolo 8, I-35131 Padova, Italy\\
$^{66}$ Centro de Investigaciones Energ\'eticas, Medioambientales y Tecnol\'ogicas (CIEMAT), Avenida Complutense 40, 28040 Madrid, Spain\\
$^{67}$ Instituto de Astrof\'isica e Ci\^encias do Espa\c{c}o, Faculdade de Ci\^encias, Universidade de Lisboa, Tapada da Ajuda, PT-1349-018 Lisboa, Portugal\\
$^{68}$ Universidad Polit\'ecnica de Cartagena, Departamento de Electr\'onica y Tecnolog\'ia de Computadoras, 30202 Cartagena, Spain\\
$^{69}$ Infrared Processing and Analysis Center, California Institute of Technology, Pasadena, CA 91125, USA\\
$^{70}$ INAF-Osservatorio Astronomico di Brera, Via Brera 28, I-20122 Milano, Italy\\
$^{71}$ Junia, EPA department, F 59000 Lille, France\\
$^{72}$ California institute of Technology, 1200 E California Blvd, Pasadena, CA 91125, USA}

\authorrunning{T.J. Moriya et al.}
\titlerunning{\textit{Euclid}: Discovering PISNe with the Deep Survey}

   \date{Received April 19, 2022; accepted August 24, 2022}

 
  \abstract{
  Pair-instability supernovae are theorized supernovae that have not yet been observationally confirmed. They are predicted to exist in low-metallicity environments. Because overall metallicity becomes lower at higher redshifts, deep near-infrared transient surveys probing high-redshift supernovae are suitable to discover pair-instability supernovae. The \textit{Euclid} satellite, which is planned to be launched in 2023, has a near-infrared wide-field instrument that is suitable for a high-redshift supernova survey.
  The Euclid Deep Survey is planned to make regular observations of three Euclid Deep Fields ($40~\mathrm{deg^2}$ in total) spanning the \textit{Euclid}’s 6 year primary mission period. While the observations of the Euclid Deep Fields are not frequent, we show that the predicted long duration of pair-instability supernovae would allow us to search for high-redshift pair-instability supernovae with the Euclid Deep Survey. Based on the current observational plan of the \textit{Euclid} mission, we conduct survey simulations in order to estimate the expected numbers of pair-instability supernova discoveries. We find that up to several hundred pair-instability supernovae at $z\lesssim 3.5$ can be discovered within the Euclid Deep Survey. We also show that pair-instability supernova candidates can be efficiently identified by their duration and color that can be determined with the current Euclid Deep Survey plan. We conclude that the \textit{Euclid} mission can lead to the first confirmation 
  of pair-instability supernovae if their event rates are as high as those predicted by recent theoretical studies. We also update the expected numbers of superluminous supernova discoveries in the Euclid Deep Survey based on the latest observational plan.
  }


   \maketitle
%

\section{Introduction}\label{sec:introduction}
Optical transient surveys have revealed the existence of huge diversity in supernovae (SNe) and many new kinds of SNe that were not recognized before have been discovered \citep{inserra2019}. For example, SNe that are often more than 10 times more luminous than other SNe, known as superluminous SNe (SLSNe, \citealt{gal-yam2019} for a recent review), were discovered through unbiased optical transient surveys \citep{quimby2011}. Just after the existence of SLSNe was recognized, they were often theoretically linked to pair-instability SNe (PISNe, e.g., \citealt{smith2007,gal-yam2009}). PISNe are theorized thermonuclear explosions of very massive stars \citep{barkat1967pisn,rakavy1967pisn} having helium core mass between $\simeq 65~\Msun$ and $\simeq 135~\Msun$ \citep[e.g.,][]{heger2002}. Some PISNe are predicted to produce a large amount of radioactive \Ni, up to $\simeq 60~\Msun$, that heats SN ejecta and they can be as bright as SLSNe \citep{kasen2011,dessart2013,gilmer2017}. However, the light-curve and spectral properties of SLSNe are now found to be different from those predicted for PISNe \citep[e.g.,][]{dessart2012,nicholl2013,jerkstrand2016,inserra2017,mazzali2019}. Therefore, it is generally believed that most SLSNe are not PISNe and they are powered by a different mechanism (see \citealt{moriya2018slsn} for a review on SLSN powering mechanisms). No SNe have been confirmed to be PISNe to date, although there are several possible PISN candidates \citep[e.g.,][]{lunnan2016,terreran2017} and some possible indirect signatures of PISNe \citep[e.g.,][]{aoki2014,isobe2021}. Confirming the existence of PISNe is a critical test of stellar evolution theory \citep{langer2012}. PISNe can play an important role in chemical evolution \citep[e.g.,][]{nomoto2013}. PISNe are also essential components for the proper understanding of black hole mass distributions observed by gravitational waves \citep[e.g.,][]{marchant2016,farmer2019}.

    \begin{table}
\caption{List of SN templates used in this study.}
\label{tab:models}      
\centering          
\begin{tabular}{cccc} 
\hline\hline       
Name & Explosion & Progenitor & Mass \\ 
\hline                    
R250\tablefootmark{a} & PISN & RSG & 250~\Msun \\
R225\tablefootmark{a} & PISN & RSG & 225~\Msun \\
R200\tablefootmark{a} & PISN & RSG & 200~\Msun \\
R175\tablefootmark{a} & PISN & RSG & 175~\Msun \\
R150\tablefootmark{a} & PISN & RSG & 150~\Msun \\
He130\tablefootmark{a} & PISN & WR & 130~\Msun \\
He120\tablefootmark{a} & PISN & WR & 120~\Msun \\
He110\tablefootmark{a} & PISN & WR & 110~\Msun \\
He100\tablefootmark{a} & PISN & WR & 100~\Msun \\
He090\tablefootmark{a} & PISN & WR & 90~\Msun \\
SLSN\tablefootmark{b} & SLSN & - & -  \\
\hline                  
\end{tabular}
\tablefoot{
\tablefoottext{a}{\citet{kasen2011}}
\tablefoottext{b}{\citet{moriya2021roman}}
}
\end{table}

   \begin{figure}
   \centering
   \includegraphics[width=\columnwidth]{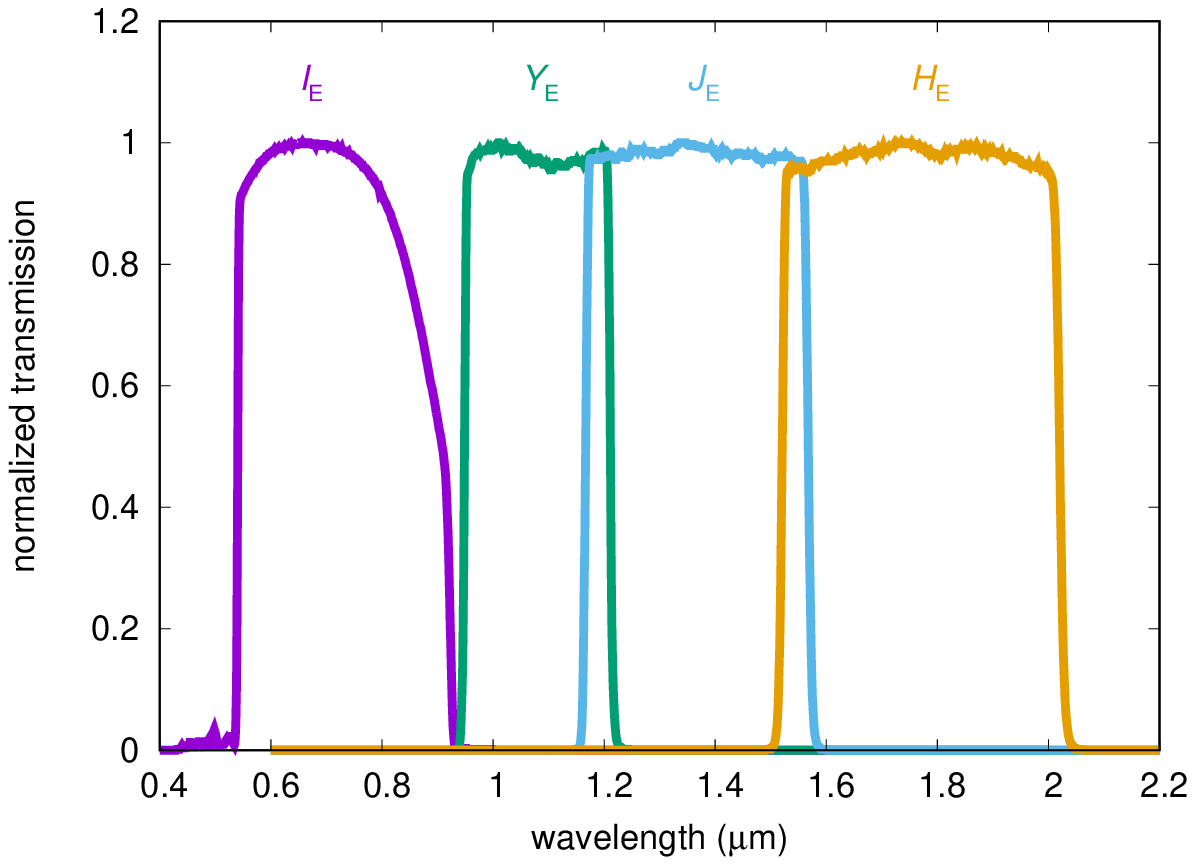}
   \caption{Normalized \textit{Euclid} filter transmission (\citealt{scaramella2021}; \citealt{schimer2022}).}
              \label{fig:filters}%
    \end{figure}
    
   \begin{figure}
   \centering
   \includegraphics[width=\columnwidth]{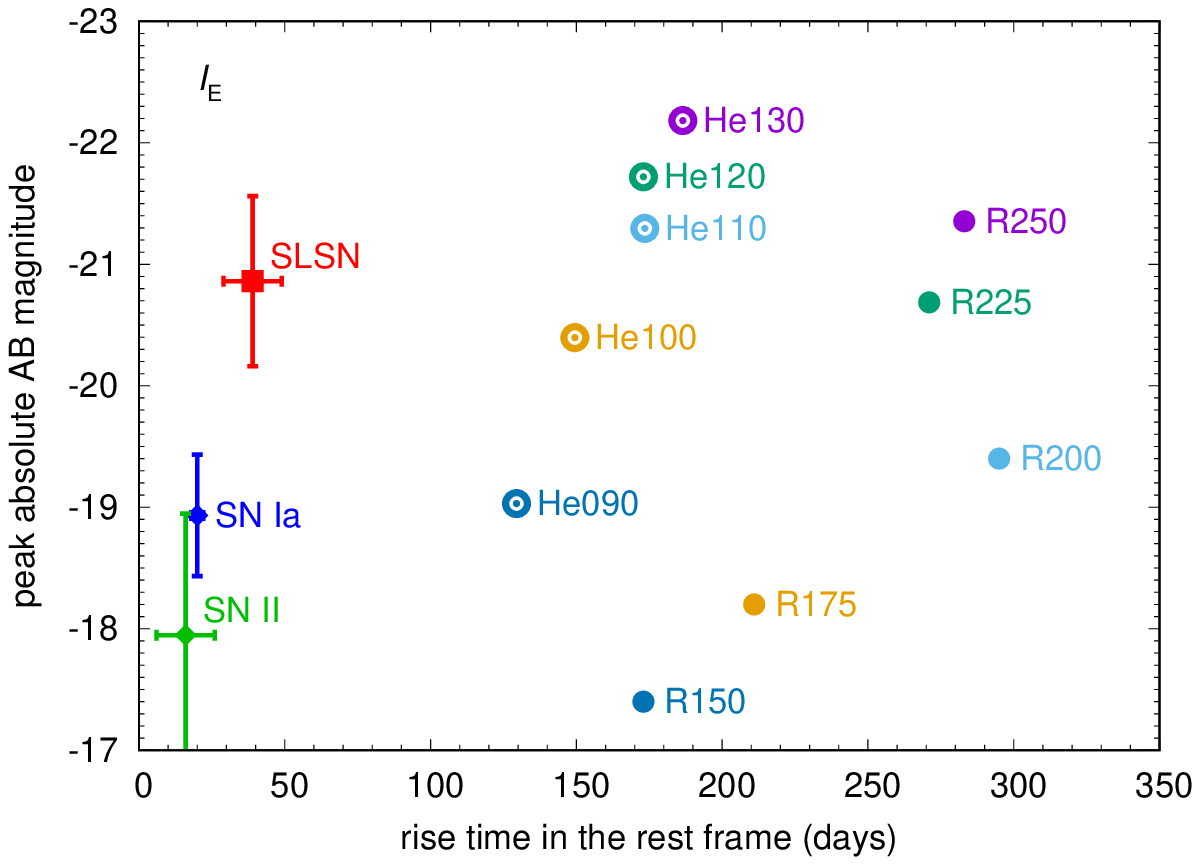}
   \caption{
   Rise times and peak absolute magnitudes of the PISN and SLSN models in the rest frame in the  $I_{\scriptscriptstyle\rm E}$ band. The PISN models (R250, R225, R200, R175, R150, He130, He120, He110, He100, and He090) are from \citet{kasen2011} and the SLSN template is from \citet{moriya2021roman}. We also show rise times and peak absolute magnitudes of the SN~Ia template from \citet{hsiao2007} and the SN~II Nugent template for comparison (see Sect.~\ref{sec:photometryscreening} for details). The error bars in the SLSN, SN~Ia, and SN~II templates show a typical dispersion \citep[e.g.,][]{richardson2014,miller2020,gonzalez-gaitan2015}.
   }
              \label{fig:risepeak}%
    \end{figure}
    
   \begin{figure}
   \centering
   \includegraphics[width=\columnwidth]{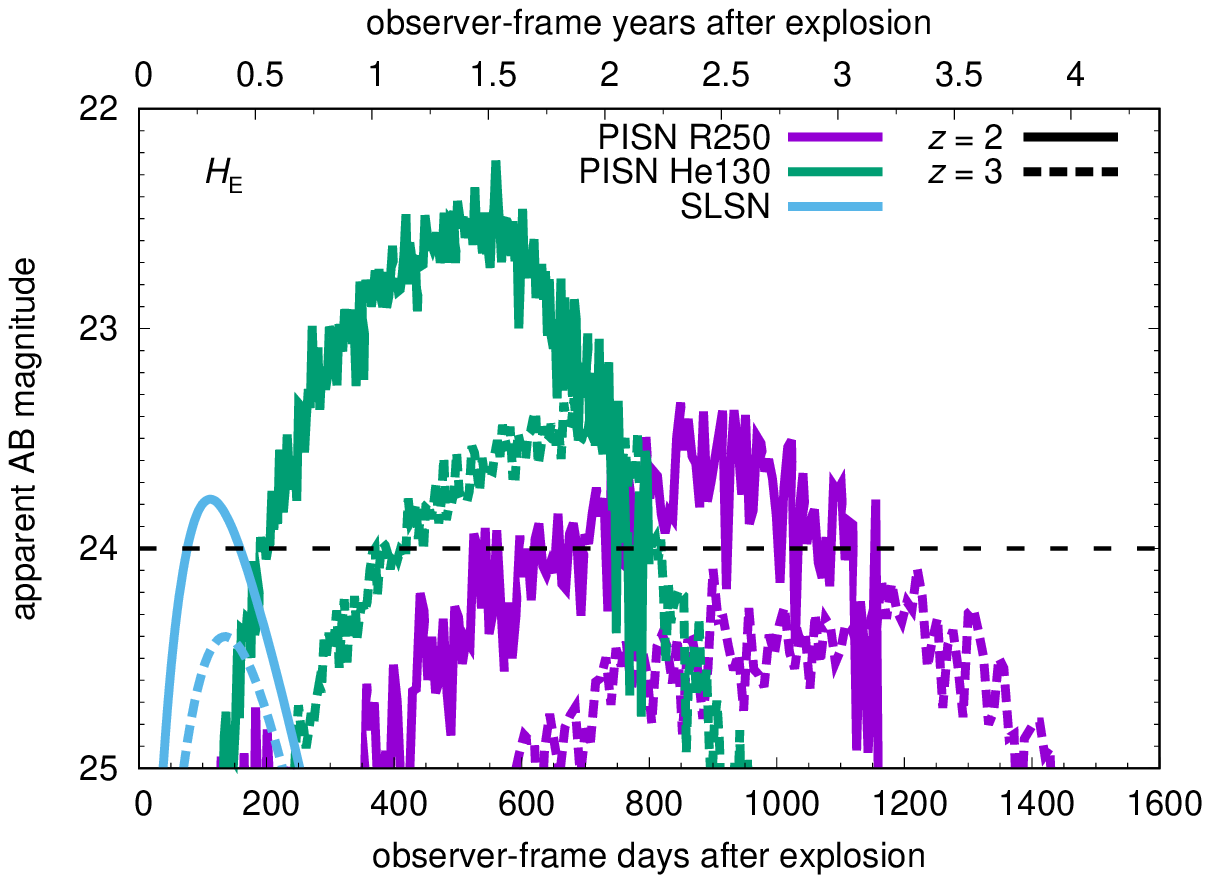}
   \caption{$H_{\scriptscriptstyle\rm E}$-band light curves of the R250 PISN model, the He130 PISN model, and the SLSN template at $z=2$ and 3. The horizontal dashed line at $H_{\scriptscriptstyle\rm E}=24.0~\mathrm{mag}$ shows the $5\sigma$ limiting magnitude of each visit planned in the EDS observations.
   }
              \label{fig:lightcurveexample}%
    \end{figure}

   \begin{figure*}
   \centering
   \includegraphics[width=\columnwidth]{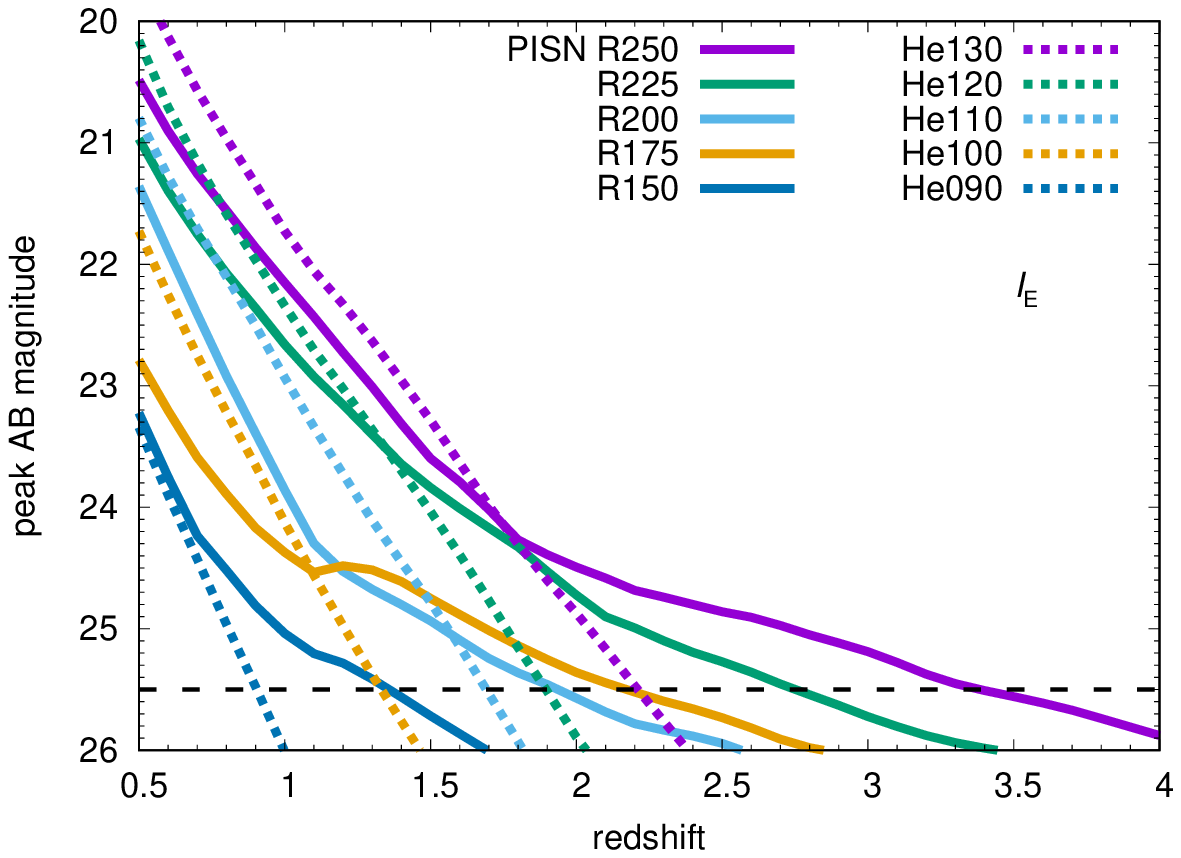}
   \includegraphics[width=\columnwidth]{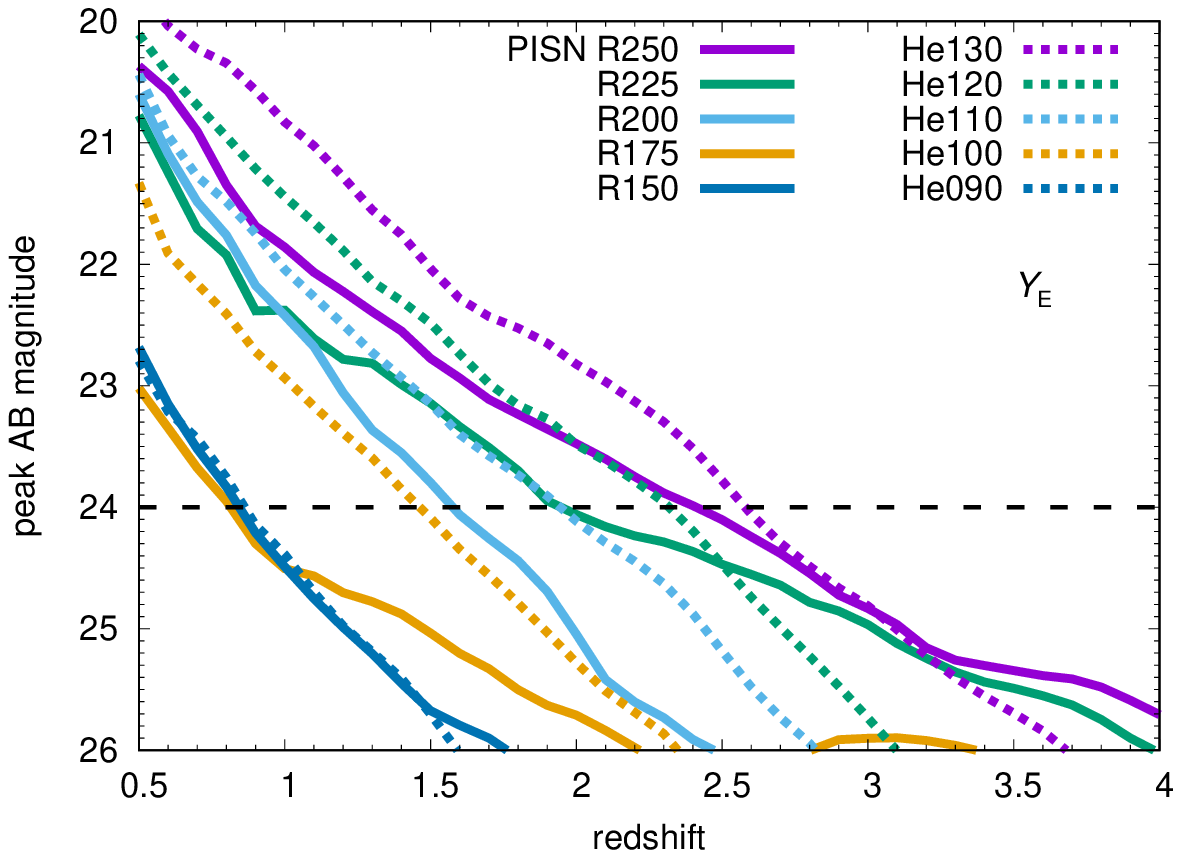} \\
   \includegraphics[width=\columnwidth]{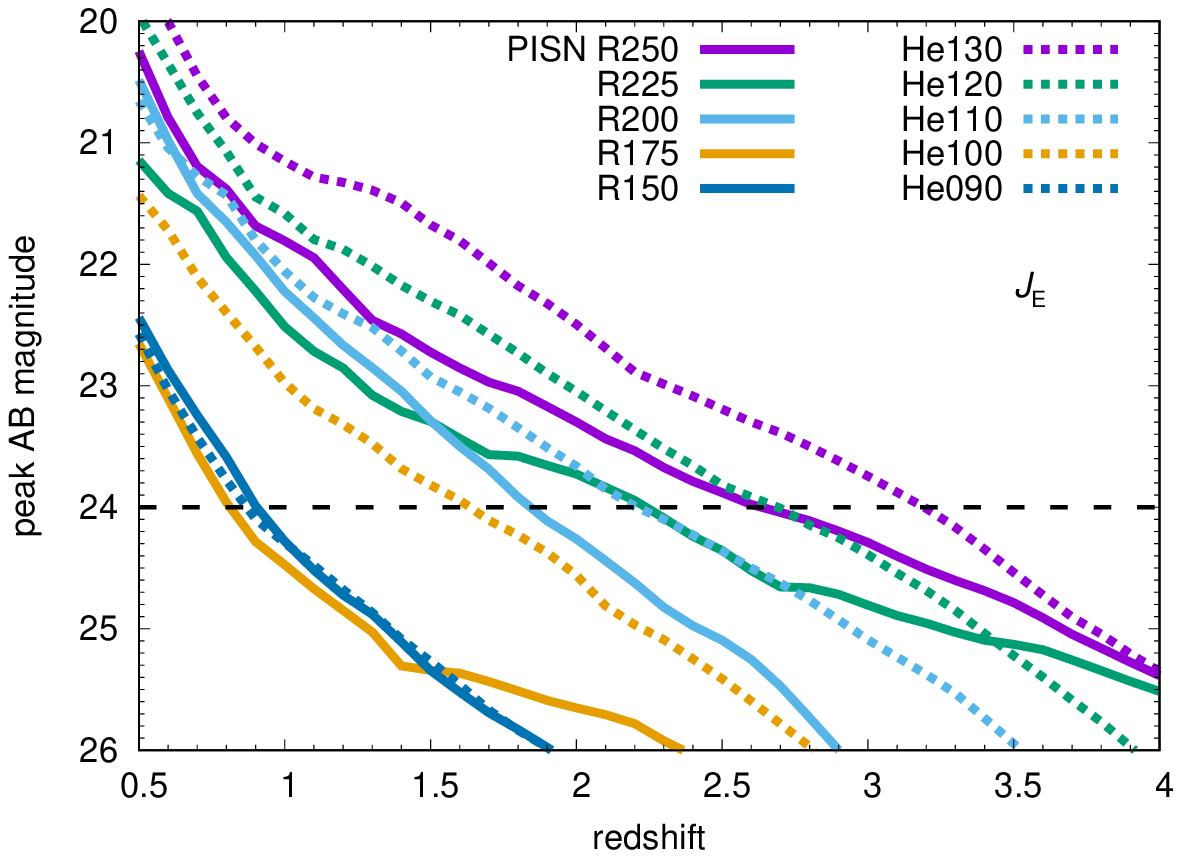}
   \includegraphics[width=\columnwidth]{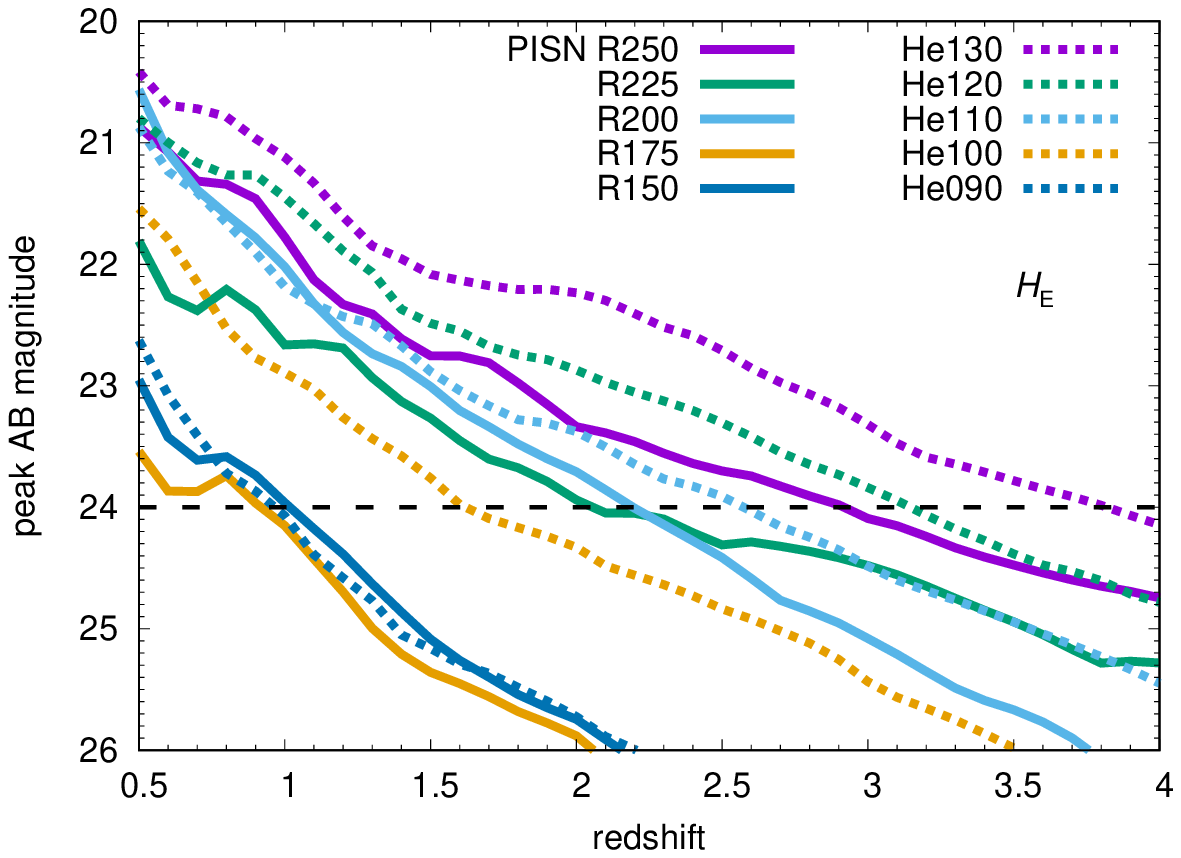}
   \caption{
   Peak magnitudes of the PISN models. The point-source $5\sigma$ limiting magnitudes of each visit adopted in our survey simulations (25.5~mag for the  $I_{\scriptscriptstyle\rm E}$ band and 24.0~mag for the $Y_{\scriptscriptstyle\rm E}$, $J_{\scriptscriptstyle\rm E}$, and $H_{\scriptscriptstyle\rm E}$ bands in AB magnitudes) are shown with the dashed lines. As discussed in Section~\ref{sec:euclidsurveyplan}, \textit{Euclid} may actually have slightly better limiting magnitudes.
   }
              \label{fig:pisnpeakmag}%
    \end{figure*}

PISNe are believed to appear in low-metallicity environments below $\mathrm{Z_\odot}/3$ (\citealt{langer2007}, but see also \citealt{georgy2017}). Thus, they may preferentially appear at high redshifts where metallicity is low \citep[e.g.,][]{chruslinska2019}. However, it is difficult to discover high-redshift PISNe with optical transient surveys as they become faint in the optical wavelength range because of redshift. For example, a deep optical transient survey with Subaru/Hyper Suprime-Cam (HSC) carried out a search for high-redshift PISN candidates, but no candidates have been found so far \citep{moriya2021pisnrate}. The future \textit{Vera C. Rubin} Observatory/Legacy Survey of Space and Time (LSST) may be able to discover PISNe depending on its survey strategy that is not determined yet \citep[e.g.,][]{dubuisson2020}. A deep transient survey in the near-infrared (NIR) is observationally better suited 
to discover PISNe appearing at high redshifts \citep{scannapieco2005,pan2012,whalen2013,moriya2019wfirst,moriya2021roman,wong2019,regos2020}.

There are several wide-field NIR instruments planned in 2020s such as the Wide Field Instrument on \textit{Nancy Grace Roman} Space Telescope from 2026 \citep{spergel2015} and the ULTIMATE Wide Field Instrument on Subaru from 2027\footnote{\url{https://ultimate.naoj.org/}}. Among them, \textit{Euclid} is planned to be launched in 2023 \citep{laureijs2011}. It has wide-field ($0.53~\mathrm{deg^2}$) imagers, one in the visual range (with a passband that we denote $I_{\scriptscriptstyle\rm E}$) and the other in the NIR range with three filters ($Y_{\scriptscriptstyle\rm E}$, $J_{\scriptscriptstyle\rm E}$, and $H_{\scriptscriptstyle\rm E}$). The \textit{Euclid} mission plans wide ($15\,000~\mathrm{deg^2}$) and deep ($40~\mathrm{deg^2}$) surveys in the four filters \citep{scaramella2021}. As introduced below, the Euclid Deep Survey (EDS) covering $40~\mathrm{deg^2}$ in total is planned to visit the survey fields regularly during the six-year primary survey period of the \textit{Euclid} mission \citep{laureijs2011,scaramella2021}. Although the current EDS plan is not suitable for searching for short-duration Type~Ia SNe (SNe~Ia, \citealt{astier2014}), \citet{inserra2018} previously showed that the regular observations planned in EDS allow us to search for high-redshift SLSNe up to $z\simeq 3.5$.

High-redshift PISNe are expected to be bright in the NIR and have long duration (Sect.~\ref{sec:pisntemplate}). Therefore, it is possible that high-redshift PISNe can be identified even with the long cadence observations planned in EDS (Sect.~\ref{sec:euclidsurveyplan}). In this paper, we perform transient survey simulations in order to quantify the expected numbers of PISN discoveries in EDS. We also investigate the expected numbers of SLSN discoveries in EDS, because the EDS observational plan has been updated since the publication of \citet{inserra2018}.

The rest of this paper is organized as follows. We first introduce the light curve templates of PISNe and SLSNe we use in our survey simulations in Sect.~\ref{sec:sntemplates}. We then introduce the assumptions in our survey simulations in Sect.~\ref{sec:surveysimulations}. The survey simulation results are presented in Sect.~\ref{sec:simulationresults}. We discuss how to identify high-redshift PISN and SLSN candidates efficiently with photometry in Sect.~\ref{sec:photometryscreening}. Section~\ref{sec:conclusions} concludes this paper. All magnitudes presented in this paper are in the AB system. We adopt the $\Lambda$CDM cosmology with $H_0=70~\mathrm{km~s^{-1}~Mpc^{-1}}$, $\Omega_\Lambda = 0.7$, and $\Omega_\mathrm{m} =0.3$.

\section{SN templates}\label{sec:sntemplates}
SN templates are required to estimate the observational properties at high redshifts. The SN templates we use in this study is introduced in this section. They are summarized in Table~\ref{tab:models}.

\subsection{PISNe}\label{sec:pisntemplate}
We use the synthetic spectral evolution of PISNe obtained by \citet{kasen2011}, who took the PISN explosion models from \citet{heger2002} and calculated their spectral evolution by performing radiative transfer calculations. \citet{kasen2011} calculated PISN properties for red supergiant (RSG), blue supergiant (BSG), and Wolf-Rayet (WR) progenitors. The RSG and BSG PISN progenitors have hydrogen-rich envelopes, while WR PISN progenitors are hydrogen-free. The BSG PISN progenitors are expected to exist mainly at zero metallicity \citep{heger2002}. The highest redshift PISNe that can be discovered by the \textit{Euclid} survey, which is determined by the maximum expected luminosity of PISNe, is $z\simeq 3.5$ as we show later and most of them are not likely to have zero metallicity. Thus, a significant fraction of PISNe discovered by \textit{Euclid} would be from RSG and WR progenitors and we adopt RSG and WR PISN models in this study. The RSG PISN models have a metallicity of $10^{-4}~\mathrm{Z_\odot}$. Although the metallicity of the host galaxies in the redshift range we can probe with EDS is higher than $10^{-4}~\mathrm{Z_\odot}$ \citep[e.g.,][]{steidel2014,chruslinska2019}, their light-curve properties are not significantly different from those with high metallicity \citep[e.g.,][]{kozyreva2014lightcurve}. We also take PISNe with WR progenitors into account in this study. The WR PISN models in \citet{kasen2011} are calculated with the zero-metallicity condition \citep[][]{heger2002}, but WR PISNe can appear with finite metallicity \citep[e.g.,][]{dubuisson2020}. The light-curve properties of zero-metallicity WR PISN models are not significantly different from those of higher metallicity WR PISNe up to the proposed PISN metallicity threshold of around $\mathrm{Z_\odot}/3$ (T.J. Moriya et al. in prep.). Thus, we adopt the WR PISN properties from the zero metallicity progenitors in this study. We note that the adopted progenitor models do not take stellar rotation into account \citep[e.g.,][]{chatzopoulos2012}. In addition, only the wind mass loss is considered.

\cite{kasen2011} provide the RSG PISN explosion models for 250~\Msun\ (R250), 225~\Msun\ (R225), 200~\Msun\ (R200), 175~\Msun\ (R175), and 150~\Msun\ (R150) stars at the zero-age main sequence and we investigate them all. We adopt the WR PISN models whose progenitor evolution is calculated from the helium zero-age main sequence with a helium core mass of 130~\Msun\ (He130), 120~\Msun\ (He120), 110~\Msun\ (He110), 100~\Msun\ (He100), and 90~\Msun\ (He090). They are the five brightest WR PISNe presented in \citet{kasen2011} whereas the WR PISN models below 90~\Msun\ are too faint to be observed in our redshift range of interest (see Sect.~\ref{sec:euclidsurveyplan}).

In order to estimate the PISN properties at high redshifts, we take the rest-frame synthetic spectra, shift them to an assumed redshift, and convolve the shifted observer-frame spectra with the \textit{Euclid} filter transmission (Fig.~\ref{fig:filters}). The rise times and peak absolute magnitudes of the PISN models in the rest frame are summarized in Fig.~\ref{fig:risepeak}. Figure~\ref{fig:lightcurveexample} shows the $H_{\scriptscriptstyle\rm E}$-band light curves of the brightest RSG PISN model (R250) and the brightest WR PISN model (He130) at $z=2$ and 3. Figure~\ref{fig:pisnpeakmag} shows the peak magnitudes of the PISN models at $0.5<z<4$ within the four \textit{Euclid} filters.

We do not consider Ly$\alpha$ absorption by the intergalactic medium (IGM) and circumgalactic medium (CGM) in this study. The IGM and CGM absorption affects the brightness in the  $I_{\scriptscriptstyle\rm E}$ band for SNe mainly at $z\gtrsim 3$. PISNe at $z\gtrsim 3$ are fainter than the  $I_{\scriptscriptstyle\rm E}$-band detection limit (25.5~mag) except for the R250 PISN model (Fig.~\ref{fig:pisnpeakmag}). In addition, the maximum redshift up to which we can discover R250 PISNe with the assumed cadence is estimated to be $z\simeq2.5$, as shown in the following sections. Thus, the Ly$\alpha$ absorption does not affect our conclusions in this study.

   \begin{figure}
   \centering
   \includegraphics[width=\columnwidth]{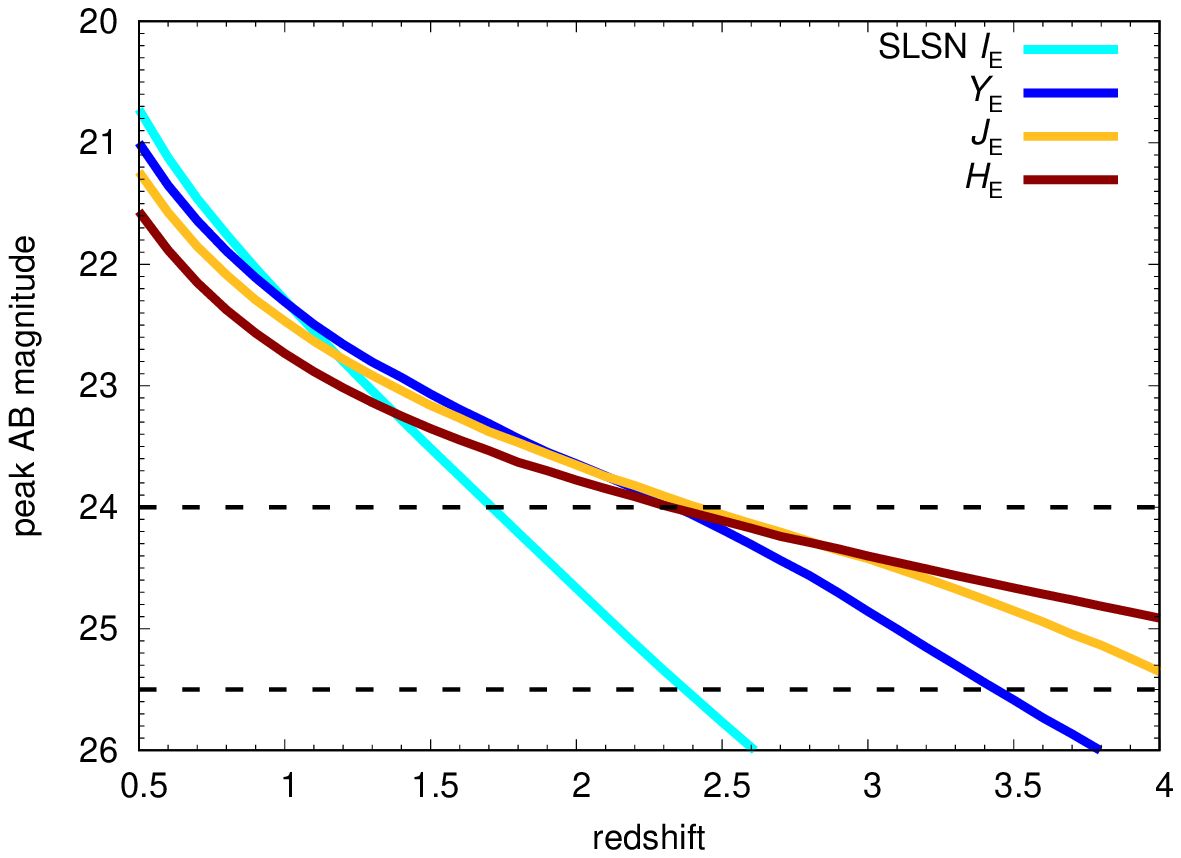}
   \caption{Peak magnitudes of the SLSN template. The limiting magnitudes of each visit by the EDS observations (25.5~mag for the  $I_{\scriptscriptstyle\rm E}$ band and 24.0~mag for the $Y_{\scriptscriptstyle\rm E}$, $J_{\scriptscriptstyle\rm E}$, and $H_{\scriptscriptstyle\rm E}$ bands in AB magnitudes) are shown with the dashed lines. The template magnitudes are randomly scattered by the standard deviation of 0.7~mag (both to brighter and fainter magnitudes) assuming the Gaussian probability distribution in the survey simulations to match the observed luminosity function as discussed in Sect.~\ref{sec:surveysimulations}.}
              \label{fig:slsnpeakmag}%
    \end{figure}

\subsection{SLSNe}\label{sec:slsntemplate}
In this study, we only consider hydrogen-poor (Type~I) SLSNe as in our previous study \citep{inserra2018} and we refer to hydrogen-poor SLSNe simply as SLSNe. This is because the observational properties of nearby hydrogen-poor SLSNe have been well investigated and we can reliably estimate their high-redshift properties. Contrarily, hydrogen-rich (Type~II) SLSNe are known to exist \citep[e.g.,][]{smith2007,inserra2018hrich}, but their rest-frame ultraviolet properties have not been studied well enough to estimate their high-redshift properties. 

We use the SLSN template presented in \citet{moriya2021roman}. The SLSN spectral template is constructed by assuming a broken blackbody function as originally introduced by \citet{prajs2017}. The template spectra above 300~nm follow the blackbody function as observed in SLSNe \citep[e.g.,][]{quimby2018}. Below 300~nm, the blackbody function is truncated to match the spectral shape of the well-observed SLSN iPTF13ajg \citep{vreeswijk2014iptf13ajg}. iPTF13ajg has the typical shape of SLSN ultraviolet spectra at around peak luminosity, although some SLSNe are known to be brighter in the ultraviolet \citep{nicholl2017uv}. We only take the spectral shape of iPTF13ajg and the template peak luminosity is adjusted to match an average value. The template has a peak \textit{u}-band absolute magnitude of $-21.1~\mathrm{mag}$, which is an average SLSN peak brightness estimated by \citet{lunnan2018}. The \textit{u}-band light-curve rise time of the template is 30~days.

The rise time and peak absolute magnitude of the SLSN template in the rest frame are compared to those of the PISN models in Fig.~\ref{fig:risepeak}. The peak observer-frame magnitudes of the SLSN template at $0.5<z<4$ obtained by convolving the \textit{Euclid} filters are presented in Fig.~\ref{fig:slsnpeakmag}. In the following survey simulations, the peak SLSN magnitudes are assumed to randomly scatter from the average value by a standard deviation of 0.7~mag with a Gaussian probability distribution in order to account for the SLSN luminosity diversity \citep{lunnan2018}. While the actual luminosity function of SLSNe may not have a Gaussian distribution function \citep[e.g.,][]{decia2018,perley2020}, a Gaussian distribution with 0.7~mag standard deviation covers the wide peak luminosity range observed in SLSNe. Similarly, the rise times are independently set to have a standard deviation of 10~days with a Gaussian probability distribution, but the rise times are not allowed to be shorter than 10~days. As mentioned in Sect.~\ref{sec:pisntemplate}, no Ly$\alpha$ absorption in the IGM and CGM is taken into account because no SLSNe at $z\gtrsim3$ are brighter than the detection limit.

   \begin{figure}
   \centering
   \includegraphics[width=\columnwidth]{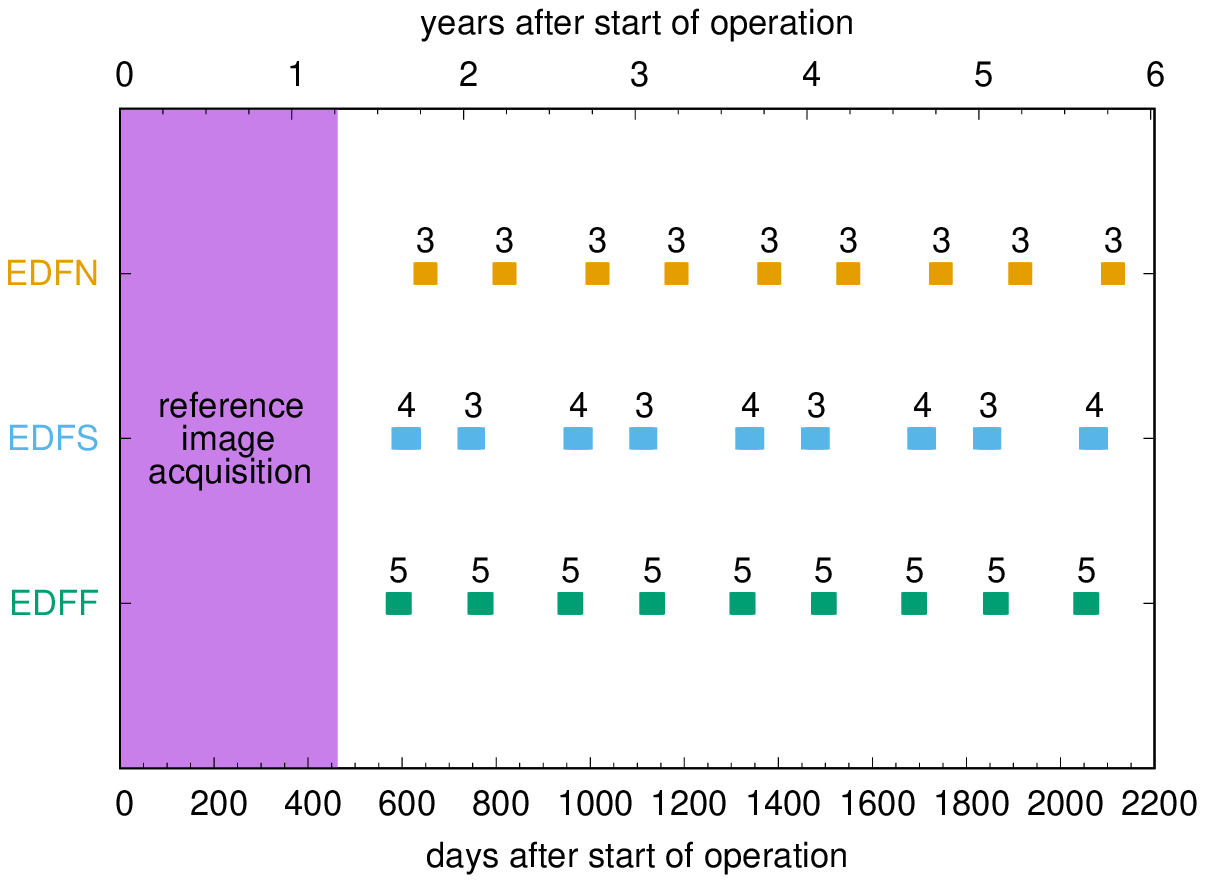}
   \caption{
   EDF observation plan of the \textit{Euclid} mission \citep{scaramella2021}. The date is from the beginning of \textit{Euclid} operation. The observational epochs at each EDF are shown with squares. The numbers above the squares show the number of visits within each epoch. For example, 5 visits for the entire $10~\mathrm{deg^2}$ of EDFF will be made at every epoch shown with squares. 
   The visits within one epoch are made within 15~days. We assume that the images taken in the first 1.25~years of the mission are used to construct the reference images and the SN search starts after 1.25~years.
   }
   \label{fig:plan}%
   \end{figure}

\section{Survey simulation setups}\label{sec:surveysimulations}
In this section, we summarize the assumptions in our survey simulations.

\subsection{Euclid Deep Fields, cadence, and sensitivity}\label{sec:euclidsurveyplan}
The EDS will cover three Euclid Deep Fields (EDFs)\footnote{\url{https://sci.esa.int/web/euclid/-/61407-the-euclid-wide-and-deep-surveys}}. Two fields are planned at the north and south ecliptic poles (EDFN and EDFS, respectively). The third field is in the Chandra Deep Field South located in Fornax (EDFF). We introduce the current observational plan at each field in this section which is also available in \citet{scaramella2021}. 

EDFN will cover $10~\mathrm{deg^2}$ and it will be visited 10 times at the beginning of the \textit{Euclid} mission for calibration purposes. Then, it will be observed regularly roughly every 6~months during the six-year primary mission period. EDFS will cover $20~\mathrm{deg^2}$ and the whole field will be visited 45 times. The first 10 visits will be made at the beginning of the mission and then the subsequent 35 visits will span 5~years. Finally, the $10~\mathrm{deg^2}$ area in EDFF will be observed 56 times in total over 6~years. 

Figure~\ref{fig:plan} summarizes the current observational plan for the EDFs. The observations at each EDF will be conducted at the epochs indicated by squares. The numbers above show the number of visits planned in each epoch. For example, EDFF is planned to have 5 visits for the entire $10~\mathrm{deg^2}$ area in every epoch shown with the squares. The visits within each epoch are made within 15~days. The brightness of high-redshift PISNe does not change significantly in 15~days (Fig.~\ref{fig:lightcurveexample}), and thus for PISNe we regard each square containing several visits as one epoch. At each visit, images in the  $I_{\scriptscriptstyle\rm E}$, $Y_{\scriptscriptstyle\rm E}$, $J_{\scriptscriptstyle\rm E}$, and $H_{\scriptscriptstyle\rm E}$ filters will be taken covering the entire survey field. The $5\sigma$ limiting magnitudes of each visit are estimated to be $I_{\scriptscriptstyle\rm E}=25.5~\mathrm{mag}$ and $Y_{\scriptscriptstyle\rm E}=J_{\scriptscriptstyle\rm E}=H_{\scriptscriptstyle\rm E}=24.0~\mathrm{mag}$ in \citet{inserra2018} and we adopted these limiting magnitudes in this study. For example, EDFF is planned to have 5 visits with the above $5\sigma$ limiting magnitudes in each epoch. We note that the recently updated estimates for the limiting magnitudes are better than those we use in this study \citep{scaramella2021,2022zndo...5836022G}. In addition, the limiting magnitudes are for each visit. Because we have multiple visits in one epoch, we can obtain slightly deeper images by stacking all visits in one epoch. Meanwhile, when we perform image subtraction for transient identification, noise will increase and the detection limits can be reduced. The noise due to image subtraction may not be significant thanks to stable \textit{Euclid} point spread functions, although the diffraction-limited NIR point spread functions are under-sampled. The actual limiting magnitudes will be revealed when the \textit{Euclid} observations start.

As shown in Fig.~\ref{fig:plan}, the EDFs will be repeatedly observed with about a half-year separation. The high-redshift PISNe are expected to be brighter than 24.0~mag in the NIR bands for a year or more (Fig.~\ref{fig:lightcurveexample}). Thus, the planned observations every half a year are suitable for searching for high-redshift PISNe. On the other hand, the typical duration of high-redshift SLSNe is not long enough for them to be captured by the half-year cadence. However, they can be still identified because there are multiple (3--5) visits made in a single epoch. We note that the EDS observation plan has been updated since the previous study of \citet{inserra2018}. Some EDFs were planned to have more frequent visits in the previous plan, which was therefore more suitable for SLSN discovery.

In order to discover transients on the sky, deep reference images are required. We assume that deep reference images are created using images taken during the first 1.25~years of the \textit{Euclid} mission as indicated in Fig.~\ref{fig:plan}. Many calibration visits are planned in the first year covering the three EDFs. The EDFs will be observed 7--13 times, depending on the field, in the first 1.25~years \citep{scaramella2021} and this number of observations is enough to create reference images of sufficient depth for subtraction.

   \begin{figure}
   \centering
   \includegraphics[width=\columnwidth]{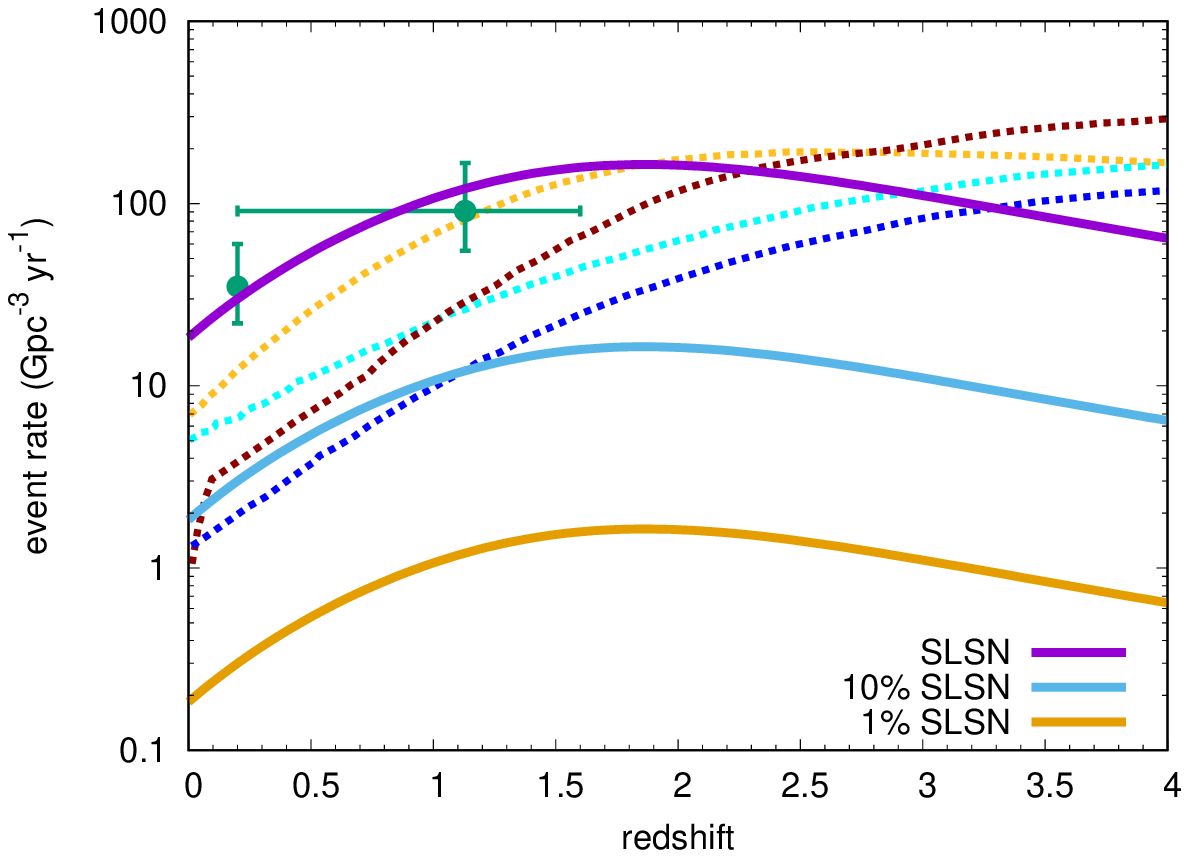}
   \caption{Event rates used in the survey simulations. The SLSN event rate is estimated by setting the SLSN event rate at $z=0.2$ to be $30~\mathrm{Gpc^{-3}~yr^{-1}}$ \citep{quimby2013,frohmaier2021} and extrapolating it following the cosmic SFR density of \citet{madau2014}. For reference, the SLSN rates constrained by observations at $z\sim 0.2$ \citep{frohmaier2021} and $z\sim 1.1$ \citep{prajs2017} are shown. We also use the 10\% and 1\% SLSN rates for PISNe. Four theoretical PISN rate estimates by \citet[][dotted lines]{briel2021} are shown for comparison. The four models have different assumptions on the cosmic SFR history.
   }
              \label{fig:slsnrate}%
    \end{figure}
    
\begin{table*}
\caption{Expected numbers of PISN and SLSN discoveries with EDS. The total discovery numbers in the three EDFs are shown.}
\label{tab:numbers}      
\centering          
\begin{tabular}{cccccccccccc} 
\hline\hline       
Model & Rate\tablefootmark{a} & Detection\tablefootmark{b} & Total & \multicolumn{7}{c}{Redshift Range} \\ 
 & & (epochs) & & 0--0.5 & 0.5--1.0 & 1.0--1.5 & 1.5--2.0 & 2.0--2.5 & 2.5--3.0 & 3.0--3.5 & 3.5--4.0\\ 
\hline                    
R250 & 1    & 2 & 145 & 4.5 & 25 & 47 & 49 & 19 & 0.030 & 0 & 0 \\  
R250 & 0.1  & 2 & 14  & 0.46 & 2.5 & 4.7 & 4.9 & 1.9 & 0.0030 & 0 & 0 \\   
R250 & 0.01 & 2 & 1.5 & 0.045 & 0.27 & 0.52 & 0.49 & 0.19 & 0.0010 & 0 & 0 \\ 
R225 & 1    & 2 & 158 & 4.4 & 25 & 47 & 53 & 28 & 0 & 0 & 0 \\ 
R225 & 0.1  & 2 & 16 & 0.46 & 2.6 & 4.8 & 5.2 & 2.7 & 0 & 0 & 0 \\ 
R225 & 0.01 & 2 & 1.6 & 0.036 & 0.25 & 0.43 & 0.54 & 0.29 & 0 & 0 & 0 \\ 
R200 & 1    & 2 & 49 & 4.4 & 23 & 21 & 0 & 0 & 0 & 0 & 0 \\ 
R200 & 0.1  & 2 & 4.9 & 0.47 & 2.4 & 2.1 & 0 & 0 & 0 & 0 & 0 \\ 
R200 & 0.01 & 2 & 0.52 & 0.056 & 0.24 & 0.23 & 0 & 0 & 0 & 0 & 0 \\ 
R175 & 1    & 2 & 7.5 & 2.9 & 4.6 & 0 & 0 & 0 & 0 & 0 & 0 \\ 
R175 & 0.1  & 2 & 0.79 & 0.29 & 0.49 & 0 & 0 & 0 & 0 & 0 & 0 \\ 
R175 & 0.01 & 2 & 0.081 & 0.038 & 0.044 & 0 & 0 & 0 & 0 & 0 & 0 \\ 
R150 & 1    & 2 & 3.2 & 1.5 & 1.7 & 0 & 0 & 0 & 0 & 0 & 0 \\ 
R150 & 0.1  & 2 & 0.32 & 0.13 & 0.19 & 0 & 0 & 0 & 0 & 0 & 0 \\ 
R150 & 0.01 & 2 & 0.033 & 0.018 & 0.015 & 0 & 0 & 0 & 0 & 0 & 0 \\ 
He130& 1    & 2 & 210 &   4.2 &  24 &  47 &  53 &  43 &  28 &  12 &   0.0060 \\ 
He130& 0.1  & 2 & 21 &   0.41 &   2.3 &   4.7 &   5.3 &   4.3 &   2.7 &   1.2 &   0 \\ 
He130& 0.01 & 2 & 2.1 &   0.040 &   0.26 &   0.46 &   0.51 &   0.45 &   0.26 &   0.12 &   0.00050 \\ 
He120& 1    & 2 & 164 &   4.3 &  21 &  47 &  51 &  36 &   3.8 &   0 &   0 \\ 
He120& 0.1  & 2 & 16 &   0.46 &   2.1 &   4.6 &   5.1 &   3.6 &   0.37 &   0 &   0 \\ 
He120& 0.01 & 2 & 1.6 &   0.048 &   0.20 &   0.43 &   0.47 &   0.38 &   0.037 &   0 &   0 \\ 
He110& 1    & 2 & 103 &   4.3 &  20 &  45 &  32 &   2.2 &   0 &   0 &   0 \\ 
He110& 0.1  & 2 & 10 &   0.43 &   2.1 &   4.5 &   3.2 &   0.24 &   0 &   0 &   0 \\ 
He110& 0.01 & 2 & 1.1 &   0.051 &   0.20 &   0.47 &   0.32 &   0.027 &   0 &   0 &   0 \\ 
He100& 1    & 2 & 25 &   4.1 &  16 &   4.7 &   0 &   0 &   0 &   0 &   0 \\ 
He100& 0.1  & 2 & 2.4 &   0.40 &   1.6 &   0.45 &   0 &   0 &   0 &   0 &   0 \\ 
He100& 0.01 & 2 & 0.24 &   0.049 &   0.16 &   0.038 &   0 &   0 &   0 &   0 &   0 \\ 
He090& 1    & 2 & 2.0 &   1.9 &   0.16 &   0 &   0 &   0 &   0 &   0 &   0 \\ 
He090& 0.1  & 2 & 0.20 &   0.18 &   0.024 &   0 &   0 &   0 &   0 &   0 &   0 \\ 
He090& 0.01 & 2 & 0.027 &   0.027 &   0 &   0 &   0 &   0 &   0 &   0 &   0 \\ 
SLSN & 1    & 2 & 28 &   2.6 &   8.5 &   8.9 &   5.1 &   2.1 &   0.75 &   0.25 &   0.096 \\ 
SLSN & 1    & 1 & 121 &   4.9 &  23 &  37 &  31 &  16 &   6.4 &   2.1 &   0.42 \\ 
\hline                  
\end{tabular}
\tablefoot{
\tablefoottext{a}{The event rate relative to the SLSN rate presented in Fig.~\ref{fig:slsnrate}.}
\tablefoottext{b}{The minimum number of detected epochs required to be regarded as a discovery.}
}
\end{table*}

   \begin{figure}
   \centering
   \includegraphics[width=\columnwidth]{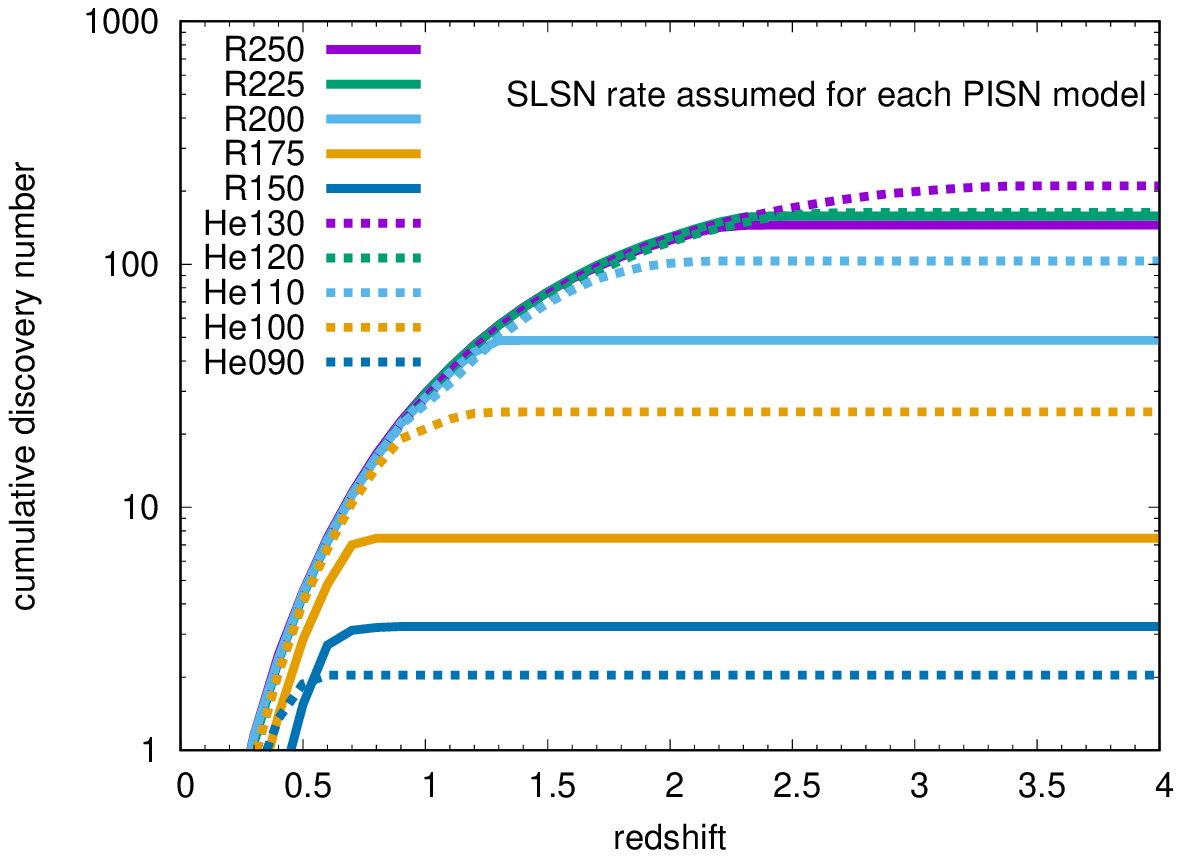}
   \includegraphics[width=\columnwidth]{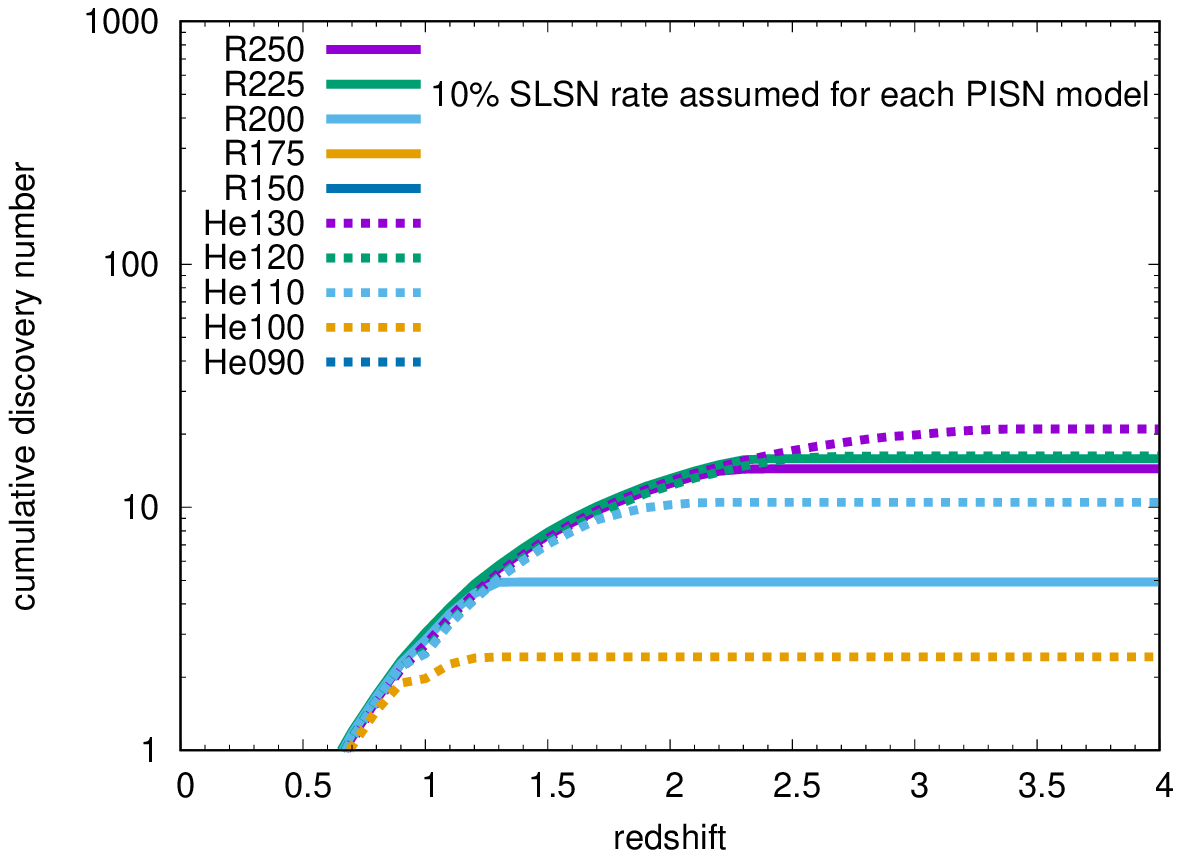} 
   \includegraphics[width=\columnwidth]{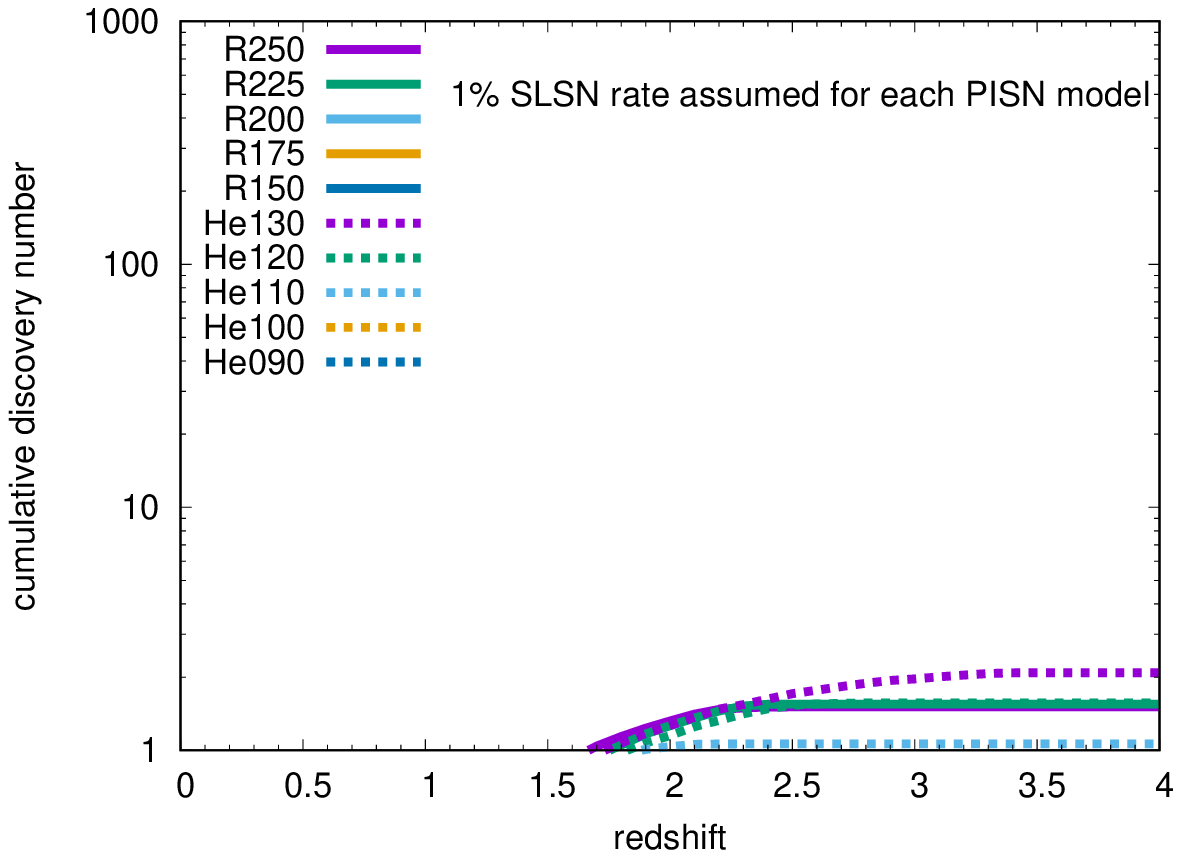}
   \caption{Cumulative redshift distributions of PISNe discovered in the survey simulations. The total discovery numbers in the three EDFs are presented. The top, middle, and bottom panels show the results for the event rates of 100\%, 10\%, and 1\% of the SLSNe rate for each PISN model.}
              \label{fig:cumulative_redshiftdistribution_pisn}%
    \end{figure}

   \begin{figure}
   \centering
   \includegraphics[width=\columnwidth]{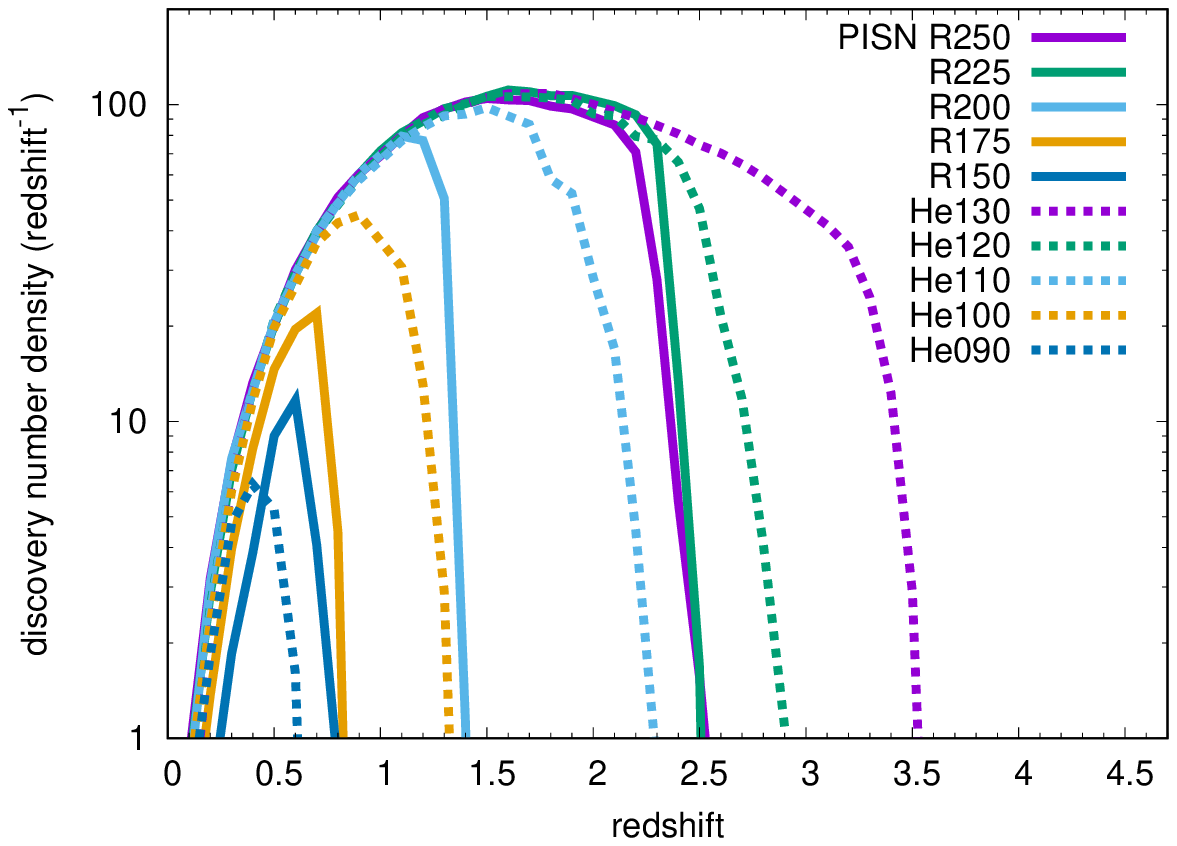}
   \includegraphics[width=\columnwidth]{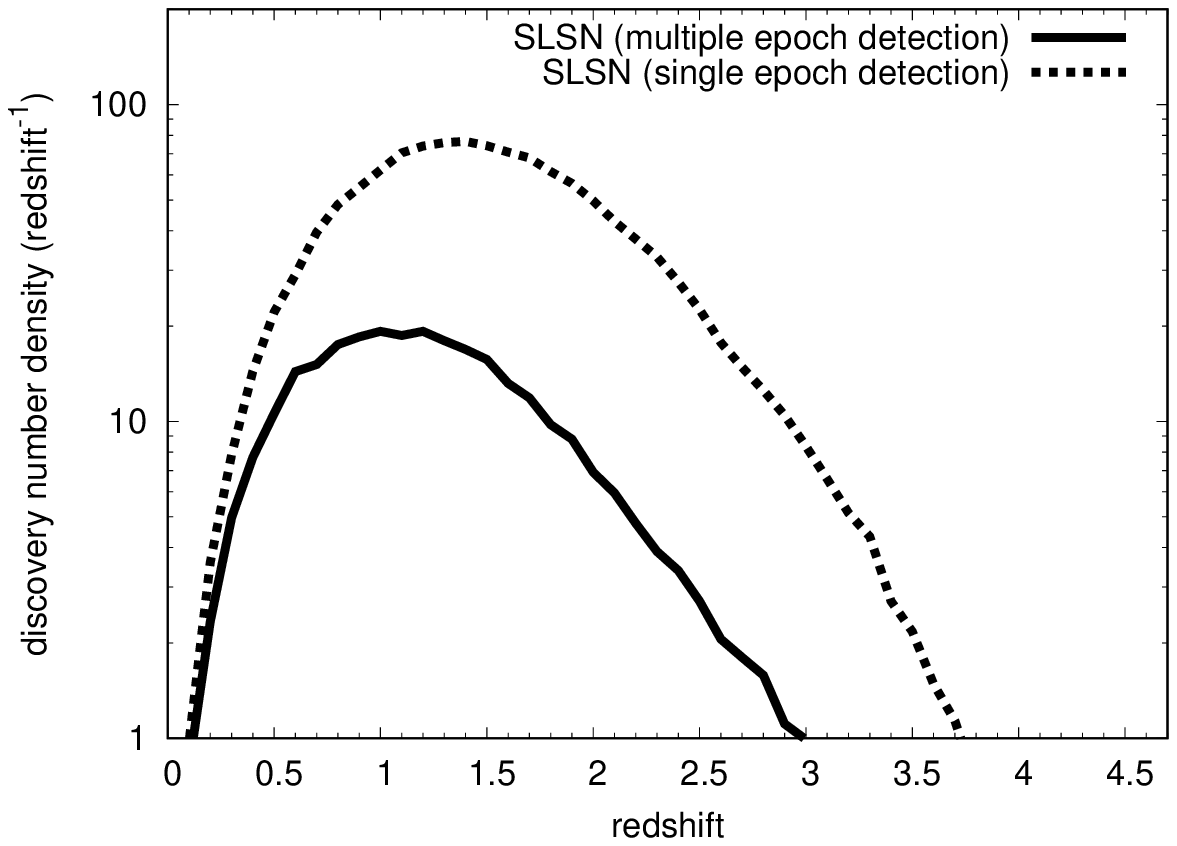} 
   \caption{
   Redshift distributions of PISNe (top panel) and SLSNe (bottom panel) discovered in the survey simulations. The total discovery numbers in the three EDFs are presented. The event rates are set to be the same as the SLSN rate in all the models in this figure, but the shape of the redshift distribution is not affected by the assumed rate.
   }
   \label{fig:redshiftdistribution}%
   \end{figure}

\subsection{Event rates}
Event rates are an essential assumption in the transient survey simulations. Because no confident PISNe have been discovered so far, the PISN event rate is unknown. Observationally, the PISN event rate is constrained to be less than the SLSN event rate \citep[e.g.,][]{nicholl2013,moriya2021pisnrate}. The SLSN event rate and its redshift evolution have been constrained by optical transient surveys \citep{quimby2013,prajs2017,frohmaier2021}. The SLSN event rate at $z\simeq 0.2$ is estimated to be $30~\mathrm{Gpc^{-3}~yr^{-1}}$ \citep{quimby2013}. The SLSN event rate up to $z\simeq 1$ is estimated to be consistent with that obtained by extrapolating the $z\simeq 0.2$ rate following the cosmic star-formation rate (SFR) density of \citet[][Fig.~\ref{fig:slsnrate}]{madau2014}. We currently have a few possible PISN candidates as discussed in Section~\ref{sec:introduction}. The total number of SLSN discoveries is currently around $100-200$ \citep[e.g.,][]{nicholl2021,chen2022a,chen2022b}. Although it is likely inappropriate to compare the numbers of the PISN candidates and the SLSN discoveries because of their different properties especially in spectral energy distributions (Section~\ref{sec:sntemplates}), the current fraction of PISN candidates to SLSNe should be around a few per cent.

Theoretical predictions for the PISN rates are also very different (see \citealt{langer2007}; \citealt{magg2016}; \citealt{takahashi2018,farmer2019,dubuisson2020,2022arXiv220409402T}). If we simply assume a Salpeter initial mass function and a PISN mass range 150--300~\Msun\ \citep{heger2002}, 1\% of core-collapse SNe (8--25~\Msun) would be PISNe. Given core-collapse SN rates of $10^{5}$--$10^{6}~\mathrm{Gpc^{-3}~yr^{-1}}$ at $z=0$--2.5 \citep{strolger2015}, the corresponding PISN rate is $10^{3}$--$10^{4}~\mathrm{Gpc^{-3}~yr^{-1}}$, which is much higher than the aforementioned SLSN rate. Recently, \citet{briel2021} provided theoretical predictions for the PISN rates. By using several cosmic SFR densities from both observations and cosmological simulations, they predicted PISN rates based on stellar evolution models from Binary Population and Spectral Synthesis (BPASS, \citealt{eldridge2017,stanway2018}). The PISN rate predictions at $z\lesssim 3.5$ are shown to be 10--100\% of the SLSN rate (Fig.~\ref{fig:slsnrate}). 

Although the PISN event rate is unknown, we need to assume some PISN event rates for our survey simulations. We think it is reasonable to assume PISN rates that are scaled with the SLSN event rate for our purpose. The SLSN event rate is constrained observationally and we do not need to rely on uncertain theoretical rate predictions in this way. A fraction of SLSNe may be PISNe even though we do not have a conclusive case. In addition, the latest theoretical study by \citet[][]{briel2021} predicts that the PISN rate at $z\lesssim 3.5$ is 10--100\% of the SLSN rate (Fig.~\ref{fig:slsnrate}). In light of the above and given the large uncertainties involved in this type of estimates, below we make three assumptions for the PISN event rate: (i) the same as the SLSN event rate, (ii) 10\% of the SLSN event rate, and (iii) 1\% of the SLSN event rate. 

NIR and radio surveys for core-collapse SNe in the local Universe have shown that a significant fraction of the events are lost from current optical surveys as they explode in very dusty environments \citep[e.g.,][]{mannucci2007,perez-torres2009,mattila2012,kool2018,jencson2019,kankare2021,fox2021}. In particular, the fraction of star-formation activity that takes place in luminous starbursting and dusty galaxies increases sharply with redshift and becomes the dominant component of star formation at $z\simeq1$. Consequently, an increasing fraction of SNe is expected to be lost from high-redshift optical searches. Up to 60\% of core-collapse SNe are likely to be lost at $z\simeq2$ \citep{mannucci2007}. Since the SLSN rates discussed in this section are based on optical surveys, the rates extrapolated to $z\simeq4$ should be taken with some caution. They can probably be considered as lower limits to the actual events rate.

   \begin{figure}
   \centering
   \includegraphics[width=\columnwidth]{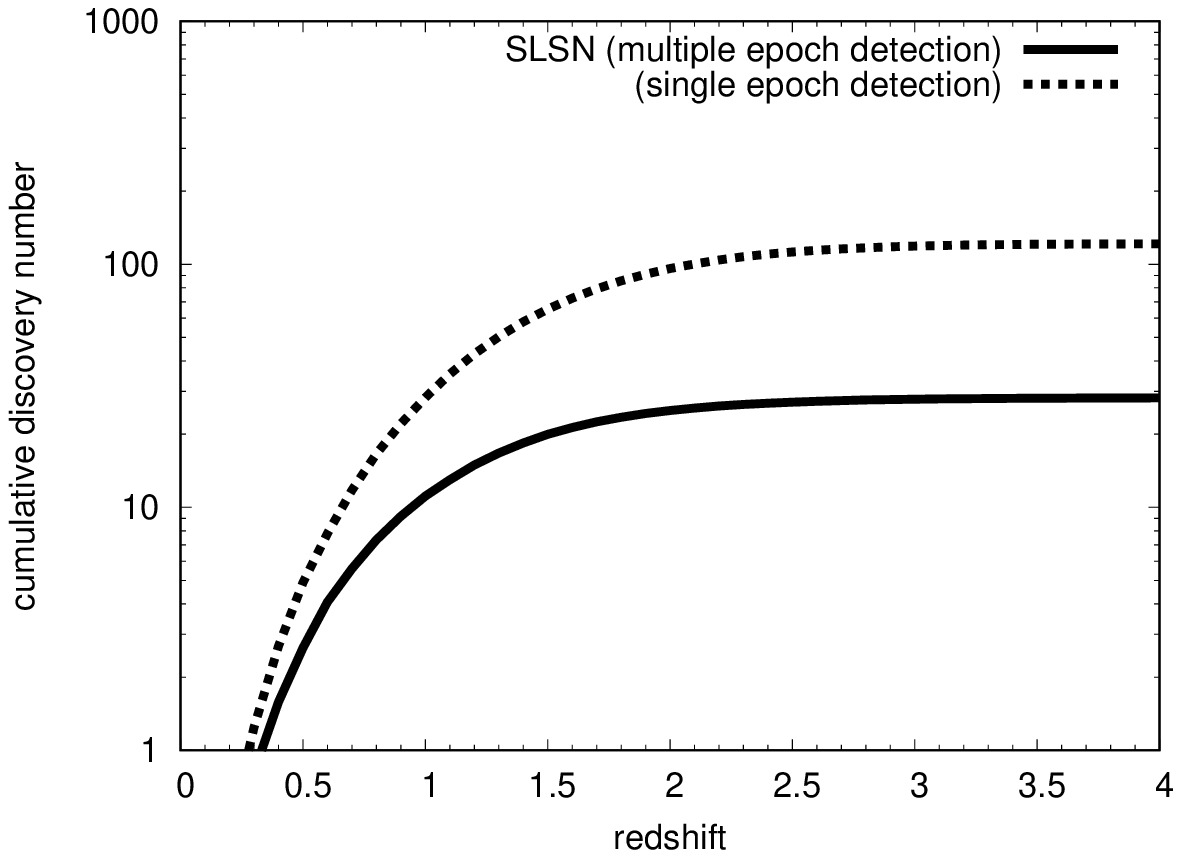}
   \caption{Cumulative redshift distributions of SLSNe discovered in the survey simulations. The total discovery numbers in the three EDFs are presented.}
              \label{fig:cumulative_redshiftdistribution_slsn}%
    \end{figure}

   \begin{figure}
   \centering
   \includegraphics[width=\columnwidth]{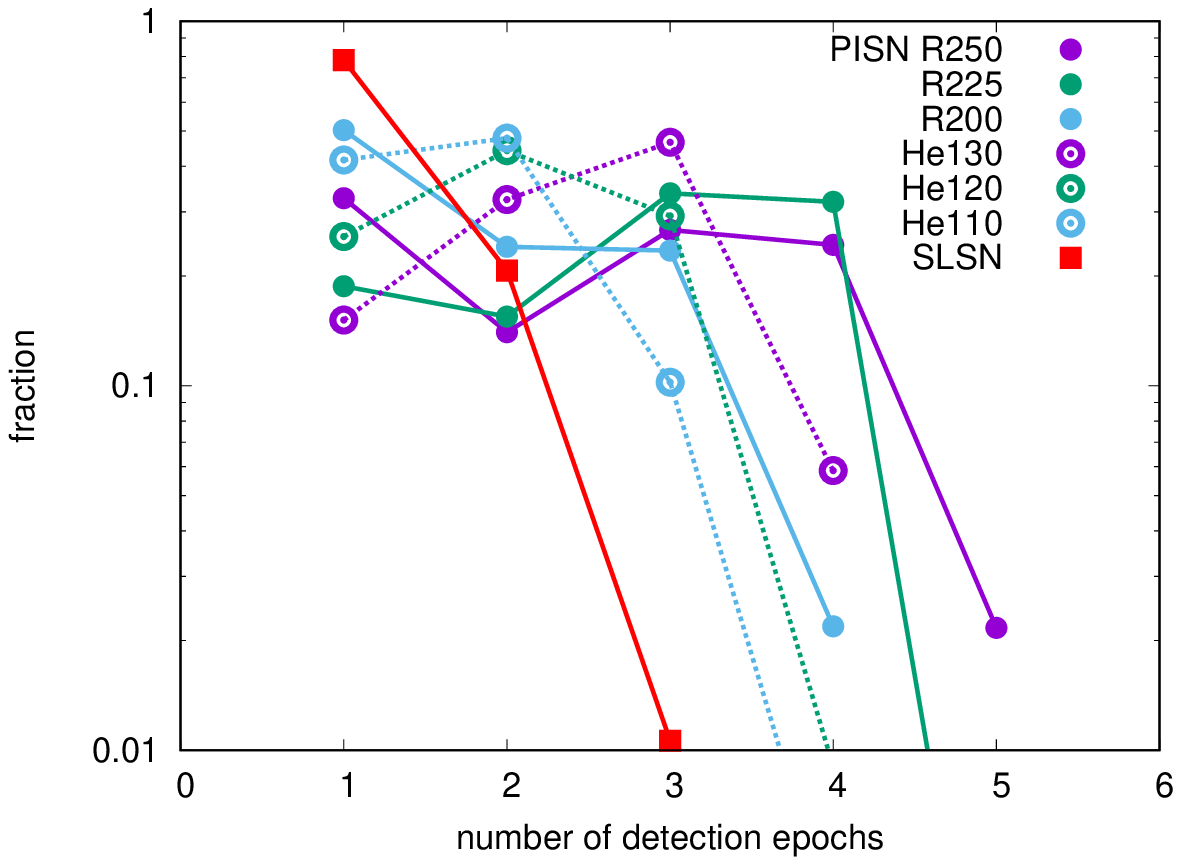}
   \caption{Distribution of the number of epochs for which luminous PISNe (R250, R225, R200, He130, He120, and He110) and SLSNe are detected with the EDFF cadence (Fig.~\ref{fig:plan}). The distribution is obtained by using the whole detected sample at all redshifts.
   }
              \label{fig:deltat_frac}%
    \end{figure}

\subsection{Discovery criteria}\label{sec:simulationsetups}
PISNe and SLSNe are assumed to appear randomly following the assumed event rates in the survey simulations. Once a PISN or SLSN appears, the expected magnitudes in each epoch are estimated for the event. If it satisfies the discovery criteria presented below, the event is regarded as a discovery.

High-redshift PISNe remain bright for a year or more (Fig.~\ref{fig:lightcurveexample}), while other common high-redshift SNe such as SNe~Ia do not remain bright for such a long time \citep{hounsell2018}, as can be seen in Fig.~\ref{fig:risepeak}. Thus, the transient duration is important information to identify rare high-redshift PISNe among many other kinds of high-redshift SNe. We, therefore, count the numbers of PISNe that become brighter than the limiting magnitude for 2~epochs or more in any band in our survey simulations. We also investigate the PISN candidate selection through their color in Sect.~\ref{sec:photometryscreening}. Even if we have a detection only in a single band, non-detection in the other bands can constrain the color of SN candidates.

High-redshift SLSNe usually have shorter durations than those of PISNe (Figs.~\ref{fig:risepeak} and \ref{fig:lightcurveexample}) and it is difficult to distinguish them from other high-redshift SNe by duration with the current EDS plan having a cadence of about a half year (Fig.~\ref{fig:plan}). However, high-redshift SLSNe can be identified among other short-duration SNe by their color even with a single epoch when they are brighter than 24~mag in the NIR bands (Sect.~\ref{sec:photometryscreening}). We count the number of SLSNe that become brighter than the detection limit at the first visit of each epoch in any band, although many other short-duration SNe will be detected with this criterion (Sect.~\ref{sec:photometryscreening}). For comparison, we also show the number of SLSNe that exceed the detection limits for two epochs.

\section{Simulation results}\label{sec:simulationresults}
Survey simulations with one set of assumptions are performed $10^3$ times and we obtain the average numbers of discoveries. The survey simulations are performed in the three EDFs independently with their areas and cadences as discussed in Section~\ref{sec:euclidsurveyplan}. However, there are no significant differences in the redshift distributions of the discovered PISNe and SLSNe among the three EDFs. Therefore, we show and discuss the combined numbers of PISN and SLSN discoveries in the three fields. The total numbers of PISN and SLSN discoveries in the three EDFs are summarized in Table~\ref{tab:numbers}.

Figure~\ref{fig:cumulative_redshiftdistribution_pisn} shows the cumulative redshift distributions of PISNe discovered in our survey simulations. We show the distributions for the three different assumptions for the event rates of the PISN models (100\%, 10\%, and 1\% of the SLSN rate), and the expected discovery numbers are roughly proportional to the assumed rate. The brightest PISN models such as R250 and He130 can be discovered with EDS if their rates are more than about 1\% of the SLSN rate. Meanwhile, the faintest PISN models in our survey simulations such as R150 and He090 can only be discovered if their event rates are as high as that of SLSNe. Figure~\ref{fig:redshiftdistribution} shows the redshift distributions of the discovered PISNe for the same event rates as the SLSN event rate. The shapes of the redshift distributions do not depend on the assumed rate. The brightest PISN model He130 can be discovered up to $z\simeq 3.5$. The brightest RSG PISN models (R250 and R225) can be discovered up to $z\simeq 2.5$. The faintest PISN models (R150 and He090) can be discovered only up to $z\simeq 0.6$--0.8 and, therefore, their expected discovery numbers are relatively small.

The expected cumulative number distributions for SLSNe are presented in Fig.~\ref{fig:cumulative_redshiftdistribution_slsn} and their number density distributions are shown in Fig.~\ref{fig:redshiftdistribution}. As we discussed previously, the duration of high-redshift SLSNe are much shorter than that of PISNe. Figure~\ref{fig:deltat_frac} shows the distribution of the number of detection epochs in our survey simulations. Only about 20\% of high-redshift SLSNe can be identified if we require them to be detected at multiple epochs. 28~SLSNe are expected to be detected at two epochs, while 121 SLSNe are discovered if SLSNe with only a single epoch detection are included. High-redshift SLSNe could be distinguished from high-redshift SNe~Ia and II through color as we discuss in Sect.~\ref{sec:photometryscreening}. In addition, we have multiple visits at epoch and the luminosity evolution in individual visits can also be used to distinguish high-redshift SLSNe from high-redshift SNe Ia and II.

Although we lose a large fraction of high-redshift SLSNe if we take only those detected at multiple epochs, we still have 28 SLSNe with multiple epoch detection. Because of the unknown PISN rates, these SLSNe could be contaminants. The color of PISNe and SLSNe can be similar and it is difficult to distinguish high-redshift PISNe and SLSNe with color alone (Sect.~\ref{sec:photometryscreening}). Because almost no high-redshift SLSNe are expected to be detected at 3 epochs and many high-redshift PISNe can be brighter than the detection limits for 3 epochs, transients with 3 epoch detection are likely to be genuine high-redshift PISN candidates.

Finally, no extinction is assumed in our survey simulations. The EDFs have little Galactic extinction, but the SNe could suffer from severe host galaxy extinction. However, PISNe and SLSNe preferentially appear in low-metallicity environments where little extinction is expected \citep{garn2010}. Thus, overall, the effect of extinction may not be significant.

   \begin{figure}[th!]
   \centering
   \includegraphics[width=\columnwidth]{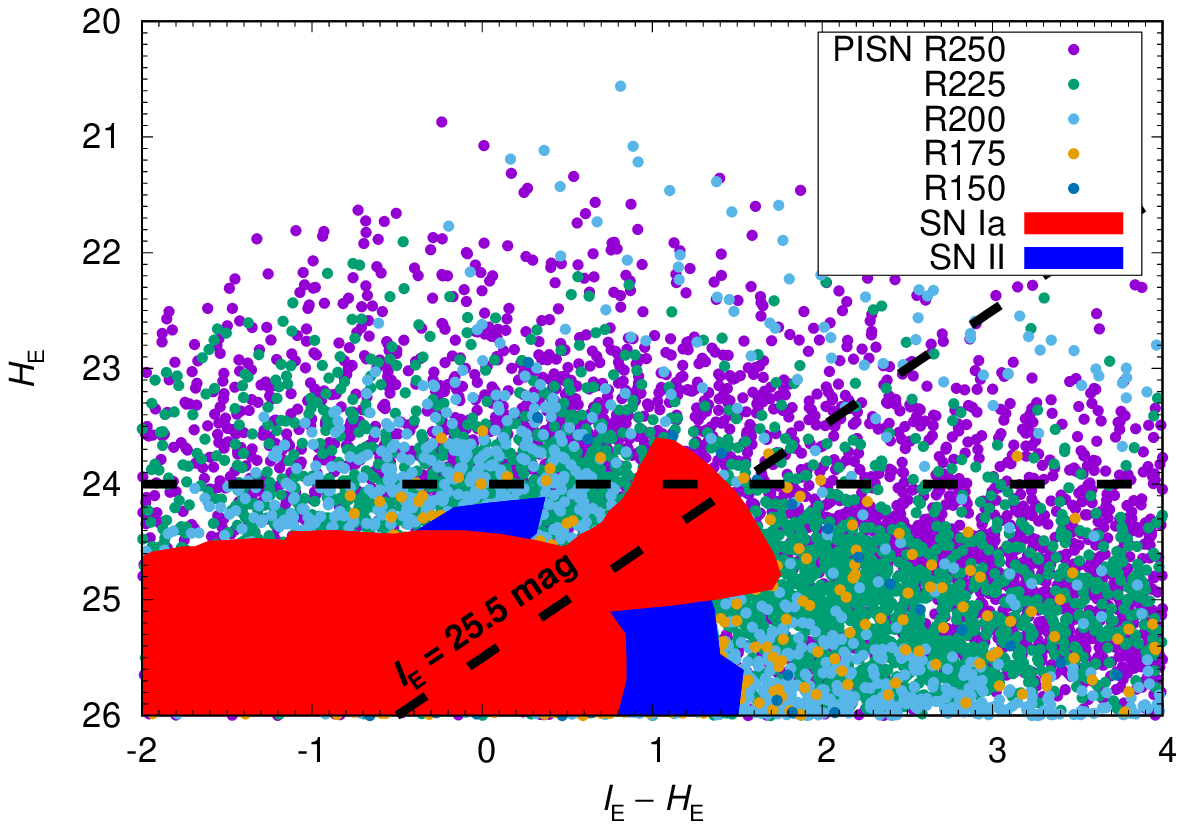}
   \includegraphics[width=\columnwidth]{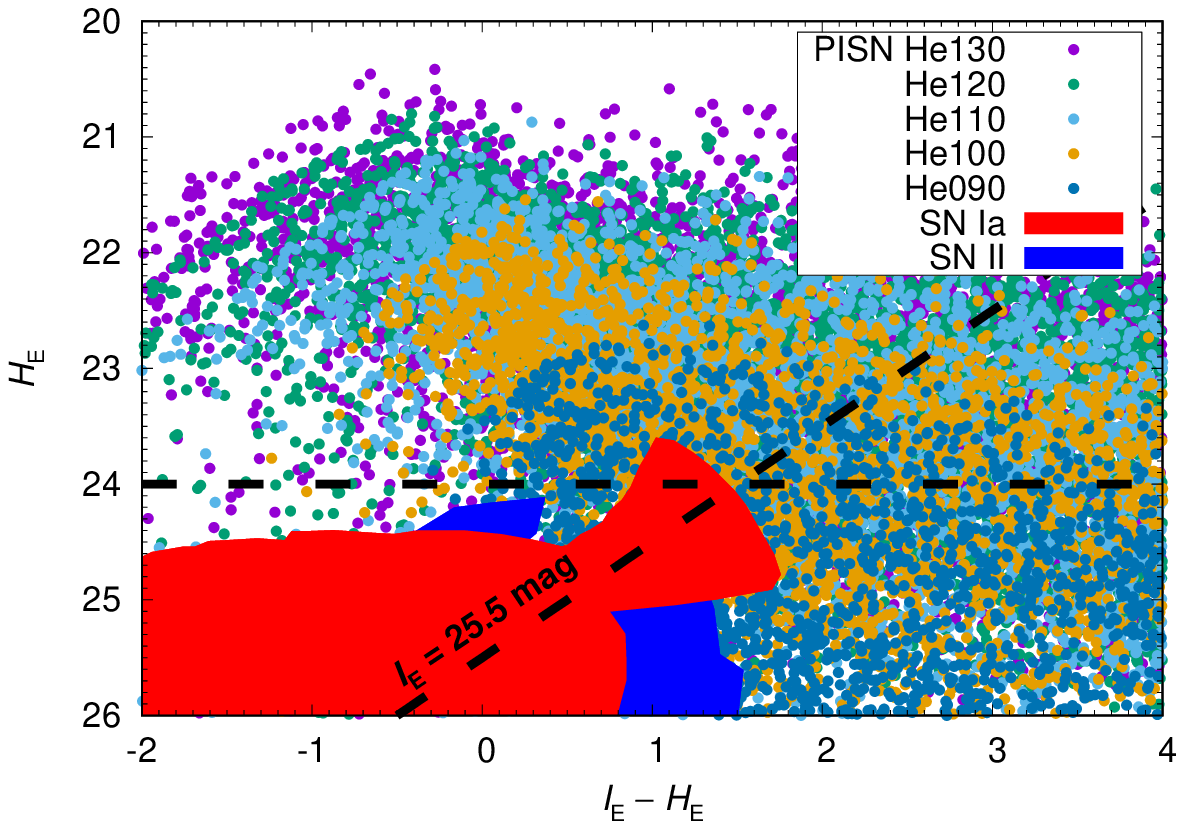}   
   \includegraphics[width=\columnwidth]{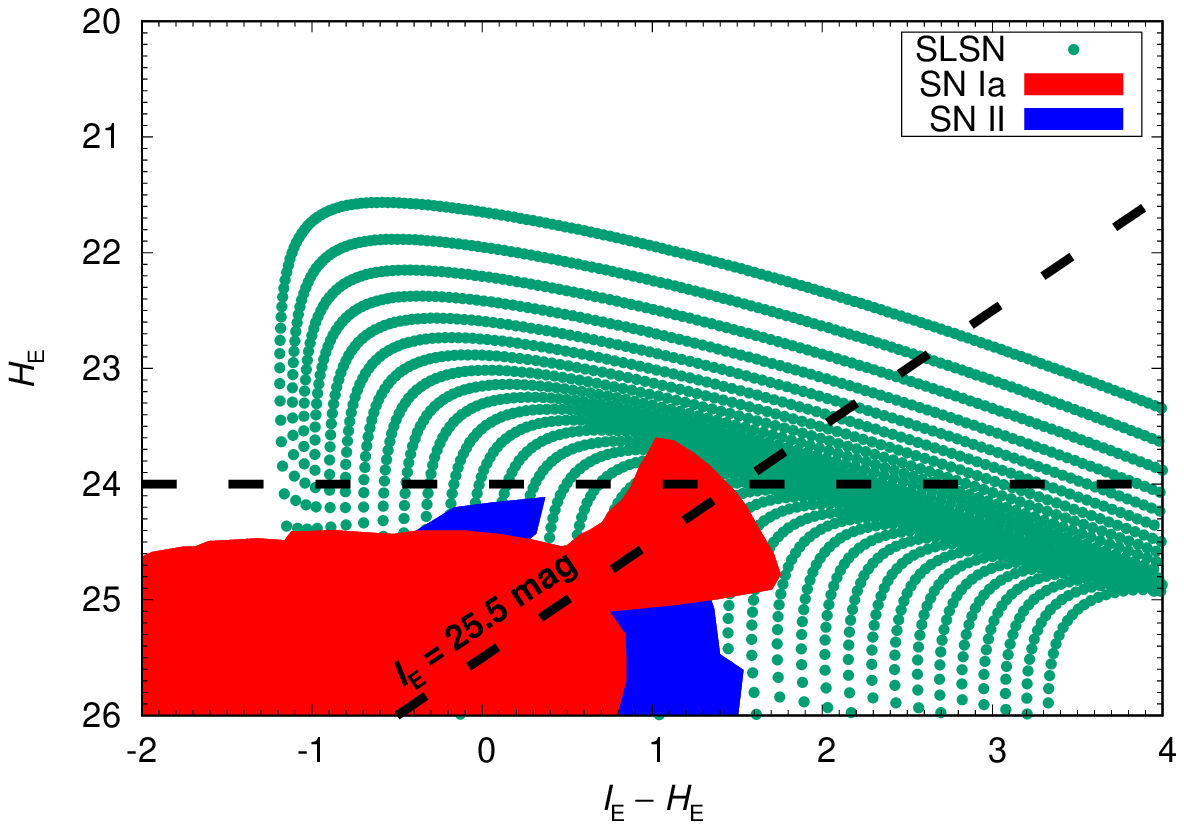}   
   \caption{Color-magnitude diagrams in the $I_{\scriptscriptstyle\rm E}$ and $H_{\scriptscriptstyle\rm E}$ bands for RSG PISNe (top), WR PISNe (middle), and SLSNe (bottom). PISNe and SLSNe at $z>0.5$ at all phases are plotted with a redshift interval of 0.1. The color evolution with time can be found in Fig.~\ref{fig:colorevolution}. The regions covered by SNe~Ia and SNe~II at $z>0.5$ are shown. No extinction is assumed. The survey limiting magnitudes for the $I_{\scriptscriptstyle\rm E}$ and $H_{\scriptscriptstyle\rm E}$ bands (25.5~mag and 24.0~mag in AB magnitudes, respectively) are indicated by the dashed lines. Most SNe~Ia and SNe~II at $z>0.5$ are too faint to be detected in the $H_{\scriptscriptstyle\rm E}$ band with the 24.0~mag limit.}
              \label{fig:cmd}%
    \end{figure}

   \begin{figure}
   \centering
   \includegraphics[width=\columnwidth]{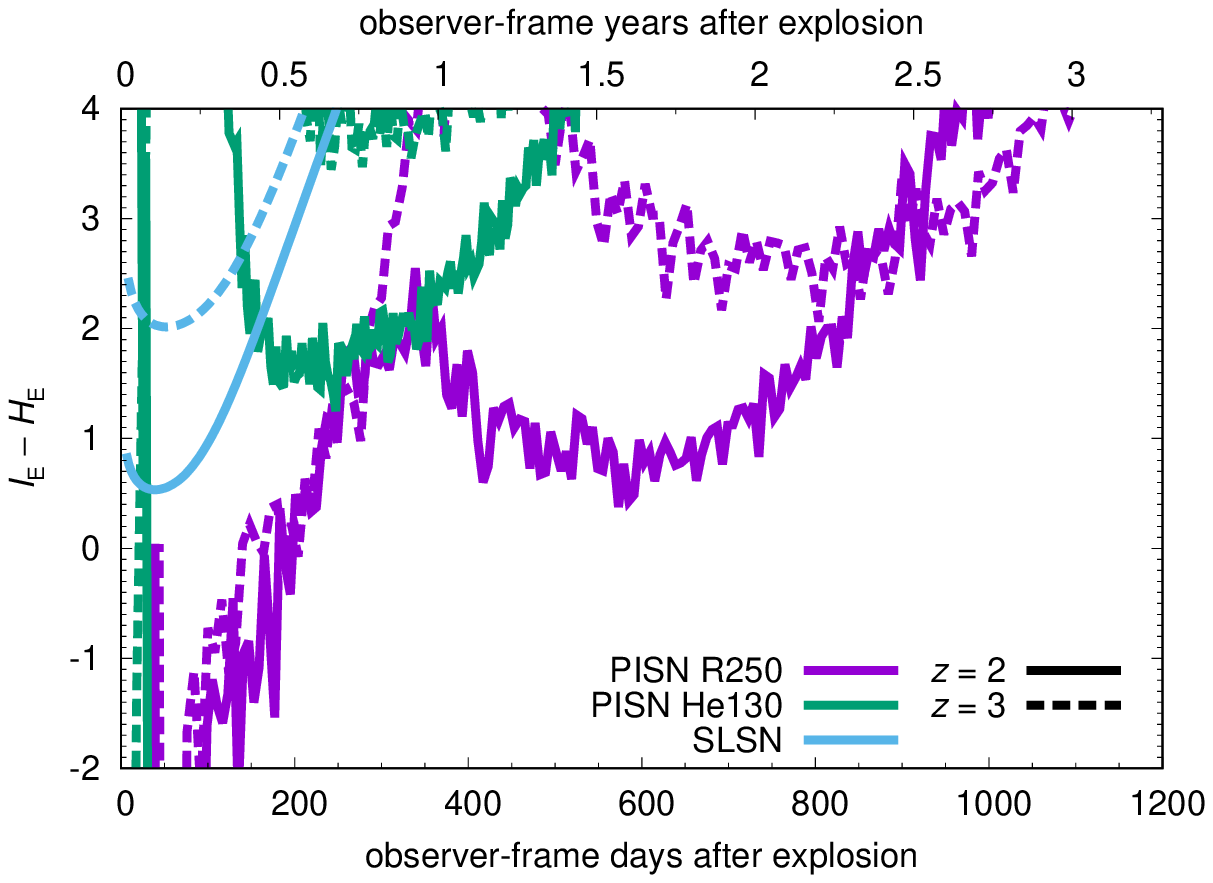}
   \caption{
   Color evolution of the PISN models and SLSN template presented in Fig.~\ref{fig:lightcurveexample}.
   }
              \label{fig:colorevolution}%
    \end{figure}

\section{PISN and SLSN candidate identification with photometry}\label{sec:photometryscreening}
Efficient identification of rare PISN and SLSN candidates among many other abundant SNe such as SNe~Ia is critical for triggering spectroscopic follow-up observations to confirm their discovery. The transient duration is essential information to identify PISN candidates especially, as discussed in the previous section. Figure~\ref{fig:deltat_frac} clearly shows that most luminous PISNe are detected at more than 2 epochs, while other SNe are usually expected to be bright for one epoch with the half-year cadence observations. Still, it is important to have as much information as possible to screen the best PISN and SLSN candidates to make the best use of follow-up resources. In this section, we discuss the color of PISNe and SLSNe, which can be used to distinguish them from other common SNe.

Figure~\ref{fig:cmd} shows the color-magnitude diagrams in the $I_{\scriptscriptstyle\rm E}$ and $H_{\scriptscriptstyle\rm E}$ bands for RSG PISNe at $z>0.5$ (top), WR PISNe at $z>0.5$ (middle), and SLSNe at $z>0.5$ (bottom). Each point shows color and magnitude for a certain epoch and all epochs in the models are presented in the figure. We also present the expected color and magnitude regions for SNe~Ia and SNe~II at $z>0.5$, assuming that those at $z<0.5$ can be discovered by optical transient surveys with, e.g., Subaru/HSC, Blanco/Dark Energy Camera, or the future \textit{Vera C. Rubin} Observatory/LSST. A few SNe per square degree at $z<0.5$ are likely to appear at each epoch \citep{yasuda2019}. The expected SN~Ia brightness is obtained by using the Hsiao template \citep{hsiao2007} with a peak \textit{B}-band absolute magnitude of $-19.26~\mathrm{mag}$ \citep{richardson2014}, which is an average absolute magnitude at peak. The SNe~II includes SNe~IIP, SNe~IIL, and SNe~IIn from the Nugent template\footnote{\url{https://c3.lbl.gov/nugent/nugent_templates.html}}. The peak absolute magnitudes for SNe~II are set as $-18~\mathrm{mag}$ in the \textit{V} band in all the subtypes. The assumed SN~II peak absolute magnitude is brighter than the average \citep{richardson2014} but the bright SNe~II are preferentially discovered at high redshifts.

Figure~\ref{fig:cmd} shows that high-redshift SNe~Ia and SNe~II can be as bright as high-redshift PISNe and SLSNe in the $I_{\scriptscriptstyle\rm E}$ band, but high-redshift SNe~Ia and SNe~II do not become as bright as high-redshift PISNe and SLSNe in the $H_{\scriptscriptstyle\rm E}$ band. Thus, the color between the $I_{\scriptscriptstyle\rm E}$ and $H_{\scriptscriptstyle\rm E}$ bands allows us to efficiently distinguish high-redshift PISNe and SLSNe from other SNe. While we do not assume any extinction in Fig.~\ref{fig:cmd}, the fact that SNe~Ia and II are detected in the $I_{\scriptscriptstyle\rm E}$ band but not in the $H_{\scriptscriptstyle\rm E}$ band indicate that their detections.

Host galaxy information of transients is also an important clue to identify their redshift information. The reference images taken at the beginning of EDS as well as deep optical images available in the EDFs would allow us to estimate host-galaxy photometric redshifts if the host galaxies are bright enough. Some host-galaxy spectra needed to determine their redshifts could be simultaneously obtained by the grisms on \textit{Euclid}. The limiting magnitude of the reference images in the $I_{\scriptscriptstyle\rm E}$ band are around 26~mag or deeper. Therefore, host galaxies with absolute magnitudes brighter than around $-19~\mathrm{mag}$ in the rest-frame ultraviolet can be identified in the reference images. Many galaxies are fainter than $-19~\mathrm{mag}$ in the ultraviolet \citep[e.g.,][]{bouwens2021} and a large fraction of PISNe could be observed as hostless. Even if we have the host galaxy photometry, the photometric redshift estimates at high redshifts are challenging \citep[e.g.,][]{2020A&A...644A..31E}. Despite of the difficulties, host galaxy information provides additional clues to identify the nature of transients. The NIR reference images will be shallower than that of $I_{\scriptscriptstyle\rm E}$, but they can trace fainter stellar masses and some host galaxies could only appear in them.

Active galactic nuclei (AGN) can be variable over long timescales and such a long-term AGN variability can contaminate our sample of high-redshift PISN and SLSN candidates. The variable component of unobscured AGN spectra at ultraviolet to optical wavelengths is estimated to be proportional to $\nu^{1/3}$, where $\nu$ denotes the frequency \citep[e.g.,][]{kokubo2014}. Thus, the color of the AGN variability does not significantly change and it remains at $I_{\scriptscriptstyle\rm E} - H_{\scriptscriptstyle\rm E} \simeq -0.34~\mathrm{mag}$. Because the color of PISNe and SLSNe changes much more with time (Fig.~\ref{fig:colorevolution}), the color evolution is essential information to exclude AGN variability. In order to exclude AGN with short-timescale variability, simultaneous short cadence optical transient surveys are helpful. The location of transients in their host galaxies is also important to exclude AGN and the high spatial resolution of \textit{Euclid} can help to constrain the nuclear vs non-nuclear location in host galaxies. Ultimately, spectroscopic confirmation of the transient would be required to confirm the classification as PISNe or SLSNe. It is possible that a part of the spectroscopic follow-up observations will be realized by the grisms on \textit{Euclid}. \textit{James Webb} Space Telescope will also be powerful for the spectroscopic follow-up observations. Photometric observations in narrower optical bands than $I_{\scriptscriptstyle\rm E}$ may also be useful to distinguish PISNe and SLSNe because of the different rest-frame ultraviolet properties expected in them \citep[e.g.,][]{dessart2012}.

\section{Conclusions}\label{sec:conclusions}
We have estimated the expected number of PISN and SLSN discoveries in EDS, in which three EDFs ($40~\mathrm{deg^2}$ in total) are planned to be observed regularly for six years during the \textit{Euclid} mission. Although the EDF observations are not frequent (Fig.~\ref{fig:plan}), the intrinsically long duration of PISNe and SLSNe allows us to discover them in the high-redshift Universe with the current survey plan. With the currently planned limiting magnitudes, it is possible to detect PISNe and SLSNe up to $z\simeq 3.5$ (Fig.~\ref{fig:pisnpeakmag}). We find that we can discover up to several hundred PISNe with EDS if the bright PISN rate is as high as the SLSN rate (Table~\ref{tab:numbers}). The expected number of PISN discoveries exceeds unity as long as the bright PISN rate is higher than 1\% of the SLSN rate. Up to 30--120~SLSNe are expected to be discovered with the same survey, depending on the discovery criteria (Table~\ref{tab:numbers}). We also show that high-redshift PISN and SLSN candidates can be efficiently identified among abundant SNe such as SNe~Ia using their duration, color, and magnitude (Figs.~\ref{fig:risepeak}, \ref{fig:deltat_frac}, and \ref{fig:cmd}).

The predicted total PISN rates are around 10--100\% of the SLSN rate at $z\lesssim 3.5$ (Fig.~\ref{fig:slsnrate}). The bright PISNe that come from progenitor masses above $\simeq 200~\Msun$ consist of about 50\% of the entire PISN population, assuming a Salpeter initial mass function and a PISN progenitor mass range of 150--300~\Msun. Even if we take 50\% of the estimated number of discoveries from our survey simulations, our expected number of PISN discoveries exceed unity when the PISN rate is higher than 10\% of the SLSN rate. Therefore, our study indicates that the \textit{Euclid} mission provides an unprecedented opportunity to finally discover PISNe if the latest PISN rate predictions are accurate. If the PISN rate is sufficiently high, it is also possible that the \textit{Euclid} mission will lead to the discovery of many PISNe with which PISNe in different environments and cosmic time can be studied.
Identification and characterization of PISNe at $z\lesssim 3.5$ by the \textit{Euclid} mission will allow us to directly confirm the existence of such objects that have never been identified with certainty before. The PISN properties can be compared with the theoretical predictions aimed at testing the theory of the evolution of massive stars in low-metallicity environments. They can also constrain massive star formation in the early Universe by providing the formation rate of the massive stars in the PISN mass range. The PISN discovery by the \textit{Euclid} mission will be an important base to discover and characterize PISNe at higher redshifts ($z\gtrsim 3.5$) which can be reached by future missions such as the \textit{Nancy Grace Roman} Space Telescope \citep[e.g.,][]{moriya2021roman,lazar2021}. Even if we do not discover any PISNe, we will be in a position to obtain a strong constraint on the PISN event rate at $z\lesssim 3.5$ that will challenge the current understanding of stellar evolution theory.

\begin{acknowledgements}
TJM thanks Mitsuru Kokubo for helpful discussions.
TJM is supported by the Grants-in-Aid for Scientific Research of the Japan Society for the Promotion of Science (JP18K13585, JP20H00174, JP21K13966, JP21H04997).
This work was supported by JSPS Core-to-Core Program (JPJSCCA20210003).
This research has made use of the SVO Filter Profile Service (\url{http://svo2.cab.inta-csic.es/theory/fps/}) supported from the Spanish MINECO through grant AYA2017-84089.
The Euclid Consortium acknowledges the European Space Agency and a number of agencies and institutes that have supported the development of \textit{Euclid}, in particular the Academy of Finland, the Agenzia Spaziale Italiana, the Belgian Science Policy, the Canadian Euclid Consortium, the French Centre National d'Etudes Spatiales, the Deutsches Zentrum f\"ur Luft- und Raumfahrt, the Danish Space Research Institute, the Funda\c{c}\~{a}o para a Ci\^{e}ncia e a Tecnologia, the Ministerio de Ciencia e Innovaci\'{o}n, the National Aeronautics and Space Administration, the National Astronomical Observatory of Japan, the Netherlandse Onderzoekschool Voor Astronomie, the Norwegian Space Agency, the Romanian Space Agency, the State Secretariat for Education, Research and Innovation (SERI) at the Swiss Space Office (SSO), and the United Kingdom Space Agency. A complete and detailed list is available on the \textit{Euclid} web site (\texttt{http://www.euclid-ec.org}).
\end{acknowledgements}

%
   \bibliographystyle{aa} 
   \bibliography{references} 
%

\end{document}